\documentclass[a4paper,fleqn,usenatbib]{aa}  

\usepackage{graphicx}	
\usepackage{amsmath}	
\usepackage{xcolor}     
\usepackage{subfig}
\usepackage{supertabular}
\usepackage{caption} 
\usepackage{times}
\usepackage{amsmath,amssymb}
\usepackage{txfonts}
\usepackage{subcaption}

\usepackage[breaklinks=true]{hyperref}
\hypersetup{
	colorlinks = true, 
	urlcolor = blue,  
	linkcolor = blue, 
	citecolor = blue 
}

\newcommand{\Msun}{M$_{\odot}$}
\newcommand{\Lsun}{L$_{\odot}$}

\newcommand{\mum}{${\mu}$m}
\newcommand{\mums}{${\mu}$m \,}

\begin{document}

   \title{An ALMA Band 7 survey of SDSS/{\it Herschel}  quasars in Stripe 82:}

   \subtitle{I. The properties of the 870 micron counterparts}

   \author{E. Hatziminaoglou\inst{1,2,3},
          H. Messias\inst{4,5},
          R. Souza\inst{6},
          A. Borkar\inst{7},
          D. Farrah\inst{8,9}
          A. Feltre\inst{10},
          G. Magdis\inst{11,12,13},\\
          L.K. Pitchford\inst{14,15},
          I. P\'erez-Fournon\inst{2,3}
          }
   \authorrunning{Hatziminaoglou et al.}
   \institute{
        ESO, Karl-Schwarzschild-Str. 2, 85748 Garching bei M\"unchen, Germany, \email{ehatzimi@eso.org}
        \and
        Instituto de Astrof\'{i}sica de Canarias, 38205 La Laguna, Tenerife, Spain 
        \and
        Departamento de Astrof\'{i}sica, Universidad de La Laguna, 38206 La Laguna, Tenerife, Spain 
        \and
        Joint ALMA Observatory, Alonso de C\'{o}rdova 3107, Vitacura 763-0355, Santiago, Chile
        \and
        European Southern Observatory, Alonso de C\'{o}rdova 3107, Vitacura, Casilla 19001, Santiago, Chile 
        \and
        Departamento de F\'{i}sica Te\'{o}rica e Experimental, Universidade Federal do Rio Grande do Norte, Campus Universit\'{a}rio, Natal RN, 59072-970, Brazil 
        \and
        Astronomical Institute, Czech Academy of Sciences, Bocn\'{i} II 1401, CZ-141 00 Prague, Czech Republic
        \and
        Institute for Astronomy, University of Hawai'i, 2680 Woodlawn Dr., Honolulu, HI, 96822, USA
        \and
        Department of Physics and Astronomy, University of Hawai'i at Mānoa, 2505 Correa Rd., Honolulu, HI, 96822, USA
        \and
        INAF - Osservatorio Astrofisico di Arcetri, Largo E. Fermi 5, I-50125, Firenze, Italy
        \and
        Cosmic Dawn Center (DAWN), Copenhagen, Denmark
        \and
        DTU Space, Technical University of Denmark, Elektrovej 327, 2800 Kgs. Lyngby, Denmark
        \and
        Niels Bohr Institute, University of Copenhagen, Jagtvej 128, 2200 Copenhagen, Denmark
        \and
        Department of Physics and Astronomy, Texas A\&M University, College Station, TX 77843-4242, USA
        \and
        George P. and Cynthia Woods Mitchell Institute for Fundamental Physics and Astronomy, Texas A\&M University, College Station, TX 77843-4242, USA
   }

   \date{Received XXX; accepted XXX}

 
  \abstract
  {Over the past 15 years, studies of quasars in the far-infrared (FIR) have reported host luminosities ranging from 10$^{12}$ to 10$^{14}$ L$_{\odot}$. These luminosities, often derived from {\it Herschel}/SPIRE photometry, suggest star formation rates (SFRs) of up to several thousand M$_{\odot}$ yr$^{-1}$, positioning them among the most luminous starburst galaxies in the Universe. However, due to the limited spatial resolution of SPIRE, there is considerable uncertainty regarding whether the FIR emission originates from the quasar itself, nearby sources at the same redshift, or unrelated sources within the SPIRE beam. To resolve this uncertainty, high-resolution observations at wavelengths close to the SPIRE coverage are required to pinpoint the true source of the FIR emission.}
  {The aim of the present work is to unambiguously identify the submillimeter (submm) counterparts of a statistical sample of FIR-bright SDSS quasars and estimate the real multiplicity rates among these systems. We study the evolution of the incidence of multiplicities with redshift, FIR properties and ``balnicity''. Based on these multiplicities, we assess the importance of mergers as triggers for concomitant accretion onto supermassive black holes (SMBHs) and extreme star formation.}
  {We conducted ALMA Band 7 continuum observations of 152 SDSS FIR-bright quasars in Stripe 82, covering redshifts between 1 and 4, with a spatial resolution of 0.8$^{\prime\prime}$. We identified all the sources detected in the Band 7 maps at or above 5$\sigma$ and performed forced photometry on the phase centre for the few quasars that were not detected otherwise. Additionally, we examined the coarse Band 7 spectra for any serendipitous detections of CO and other transitions.}
  {We find that in approximately 60\% of all cases, the submm emission originates from a single counterpart within the SPIRE beam, centred on the optical coordinates of the quasar. The rate of multiplicity increases with redshift, rising by a factor of $\sim$2.5 between redshifts 1 and 2.5. The incidence of multiplicities is consistent among broad absorption line (BAL) quasars and non-BAL quasars. The multiplicities observed in a fraction of the sample indicate that, while mergers are known to enhance gas inflow efficiency, there must be viable alternatives for driving synchronous SMBH growth and intense star formation in isolated systems. Additionally, we report the serendipitous detection of two CO(6-5) and three CO(7-6) transitions in five quasars at redshifts between 1 and 1.4, out of the eight such transitions expected based on the spectral setup and the redshifts of the objects in the sample. Higher transitions, although expected in a fraction of the sample, are not detected, indicating that the quasars are not exciting sufficiently the gas in their hosts. Finally, we also detect a potential emission of H$_2$O, HCN (10-9) or a combination of both in the spectrum of a quasar at redshift 1.67.}
  {}


   \maketitle
%

\section{Introduction}
\label{sec:intro}
Accretion onto supermassive black holes (SMBHs) and starburst activity are known to often occur concomitantly \citep[e.g.][just to name a few]{farrah03, lutz08, hernan09, hatzimi10, santini12, drouart14, pitchford16, magliocchetti22}. Both processes draw from a common gas reservoirs, possibly fed by major-merger-induced strong inflows \citep[e.g.][]{dimatteo05, hopkins06a, hopkins06b}. And while galaxy interactions make gas available for fuelling both SMBH growth and intense star formation, the same gas may become unavailable by the ejection of energy by massive outflows driven by the accretion itself, two processes that seem to be in perpetual competition with one another. Furthermore, Active Galactic Nuclei (AGN) triggering mechanisms may change with redshift, with major mergers dominating above a redshift of $\sim$1.5 and secular mechanism being more important at lower redshifts \citep{draper12}.

Despite decades of observational and modelling efforts to understand the interplay between intense star formation and accretion onto SMBHs, the precise effects of these processes on each other and on the evolution of massive galaxies remain poorly constrained. This gap in knowledge highlights the importance of studying objects where both phenomena occur simultaneously, as they serve as key laboratories for advancing our understanding of the most fundamental aspects of galaxy evolution. 

Such objects are optically- {\it and} far-infrared-bright quasars. Indeed, a small fraction ($\sim$ 5\% - 8\%) of optically bright (SDSS\footnote{\url{https://www.sdss.org/}}) quasars (with bolometric luminosities $10^{45}$ $\le$ L$_{\rm bol} \le 10^{47}$ erg/s derived by the SDSS pipeline) show extraordinarily high far-infrared (FIR) luminosities (in the range 10$^{12}$ to 10$^{14}$ \Lsun), based on their individual (as opposed to stacked) {\it Herschel}/SPIRE fluxes at 250, 350 and 500 $\mu$m \citep{caoorjales12, pitchford16,kirkpatrick20}. 
Such high FIR luminosities are suggestive of star formation rates (SFRs) of up to a few thousand \Msun yr$^{-1}$ even after the contribution of the AGN to the FIR is accounted for, placing these systems among the most luminous starburst hosts in the Universe. These SFRs are 3 to 10 times higher than the average SFRs of the remaining ~95\% SDSS quasars (undetected in the FIR down to the {\it Herschel}/SPIRE limits), estimated by stacking {\it Herschel}/SPIRE fluxes \citep{harrison12, mullaney15, harris16}. SDSS FIR-bright quasars, which appear to be undergoing simultaneous episodes of intense star formation and SMBH growth near the Eddington limit, serve as ideal laboratories for studying the physical conditions and environments that trigger and sustain both processes in parallel.

The high SFRs observed in the hosts of bright quasars, which are thought to be quenching or even suppressing star formation \citep[e.g.][]{page12}, are surprising.
As these SFR estimates are primarily based on FIR flux measurements, often from {\it Herschel}/SPIRE, the uncertainty in identifying the correct FIR counterparts also impacts their reliability. Counterpart identification has long been a challenge in FIR surveys \citep{wang14, hurley17}, and it remains uncertain whether the SFRs derived for these quasar hosts are even associated with their hosts. Furthermore, given SPIRE's spatial resolution of 18\arcsec \, at 250 \mums \citep{griffin10}, it is likely that in some cases multiple sources contribute to the observed FIR emission. The presence of two or more sources within the SPIRE beam may suggest close pairs, and this proximity could be the trigger for these two extreme phenomena. Accurate identification of the origin of the FIR emission is, therefore, crucial.  

Interferometric observations \citep{hodge13, bussmann15, trakhtenbrot17, stach18, nguyen20} have shown a fraction of single-dish submillimetre (submm) or {\it Herschel} sources to be blends of multiple galaxies. The fraction of multiple sources varies between 30\% and 70\% depending on the sensitivity of the submm observations, the nature of the objects, the sample selection and the definition of multiplicity. Simulations, on the other hand, tend to predict a higher multiplicity fraction than what is observed on average \citep{hayward13, cowley15}, possibly due to the assumptions that go into the triggering mechanisms of submm galaxies (SMGs). 

In addition to the uncertainties on the source of the FIR emission, the source of heating of the cold dust in the hosts of AGN is also a topic that has not yet fully settled. Although star formation is considered to be the driving mechanism \citep[e.g.][]{hatzimi10, pitchford16, lamperti21}, some works indicate that, in at least some cases, the cold gas may be directly heated by the AGN \citep[e.g.][]{symeonidis17,maddox17}. A more direct evidence might come from the excitation level of the molecular gas in the hosts of those FIR-bright AGN. Low-J CO transitions are believed to trace gas within Photodissociation Regions (PDRs) heated by far UV radiation while high-J lines are tracers of X-ray-dominated regions \citep{wolfire22}, already from the CO(4-3) transition and upwards \citep{esposito24}. Nevertheless, \cite{farrah22} only found the CO(10-9) transition and above to be increasingly excited by the AGN. Along similar lines, \cite{valentino21} suggest that the AGN has only a marginal effect on the CO, at least up to CO(7-6). Earlier studies also indicate that even the higher CO transitions in nearby galaxies can be explained by PDRs \citep{rigopoulou13}.

In order to address the source of FIR emission in SDSS quasars, a pilot study of 28 FIR-bright SDSS quasars at redshifts between 2 and 4 was conducted with the Atacama Compact Array (ACA) in Band 7 \citep{hatzimi18}. The study, primarily aimed to investigate the issue of multiplicities around quasars, found that about a third of these quasars show clear evidence of secondary counterparts contributing at least 25\% to the total 870 \mums continuum flux within the SPIRE beam. For a few sources, the FIR emission was not associated with the SDSS quasar at all, but instead came from a different source. The same study suggested that Broad Absorption Line (BAL) quasars might exhibit higher multiplicity rates (57\%) compared to non-BAL quasars (24\%), indicating that BAL and non-BAL quasars may reside in different environments. However, these results are limited by low-number statistics (only seven BAL quasars among the 28 objects in the sample). 
Given the small sample size, the coarse resolution of the ACA ($\sim$4\arcsec) and the relatively narrow redshift range, the results, while suggestive, were neither conclusive nor fully representative. 

The present paper builds upon the pilot study and presents the results from submm observations of 152 FIR-bright SDSS quasars, conducted with the ALMA 12-m Array in Band 7 (870 \mum) at a resolution of 0.8\arcsec. The primary aim is to verify the findings from the previous study. Specifically, this paper focuses on the multiplicities and the serendipitous detections of CO transitions and their implications. A forthcoming paper (Hatziminaoglou et al., in preparation) revisits the Spectral Energy Distribution (SED) fitting and derived SFRs discussed in \cite{pitchford16}, and further examines the properties of the cold dust in the quasar hosts.

The structure of this paper is as follows: Sections \ref{sec:sdsssample} and \ref{sec:almadata} describe the quasar sample and the ALMA data, respectively. Section \ref{sec:counterparts} discussed the ALMA detections in the quasar fields and the findings related to multiplicities, while Sect. \ref{sec:lines} presents serendipitous detections of intermediate CO transition. The paper concludes with a discussion of the results in the context of the recent literature in Sect. \ref{sec:discuss}.

\section{ALMA SDSS quasars sample}
\label{sec:sdsssample}

The quasar sample under study is drawn from the parent sample of FIR-bright SDSS quasars in the HerMES \citep{oliver12} fields analysed in \cite{pitchford16}. It includes a total of 152 FIR-bright (S$_{250} \ge 30$ mJy) type 1 SDSS quasars from the SDSS DR7 \citep{schneider07} and DR10 \citep{paris14} ($10^{45} \le \, {\rm L_{acc}} \, \le10^{47}$ erg s$^{-1}$), in the redshift range 1 $\le z \le$ 4, lying in the {\it Herschel} Stripe 82 Survey \citep[HerS;][]{viero14}. HerS consists of 79 deg$^2$ of contiguous SPIRE imaging of the SDSS Stripe 82 and has excellent quality SPIRE 250, 350 and 500 $\mu$m fluxes, extracted following \cite{hurley17}, as well as a wealth of ancillary multi-wavelength data. These include UV data from GALEX \citep{morrisey07}, near-infrared observations from UKIDSS \citep{hewett06} and VHS \citep{mcmahon21}, mid-infrared coverage from WISE \citep{wright10} and 1.4 GHz observations with FIRST \citep{white97}. 

As the parent sample was drawn from more than a decade old SDSS data releases, the quasars in the sample were checked against the SDSS Data Release 16 (DR16) quasar catalogue, that encompasses all of the quasars from previous releases \citep[for details see][]{lyke20}. The quasar nature and the redshifts of all  
sources remain unchanged in the DR16 catalogue with respect to the previous data releases. 
The sample is the largest ever observed in the submm with a uniform resolution. To the best of our knowledge, the 152 quasars are not lensed. 

\begin{figure*}
\center
\includegraphics[width=.48\textwidth]{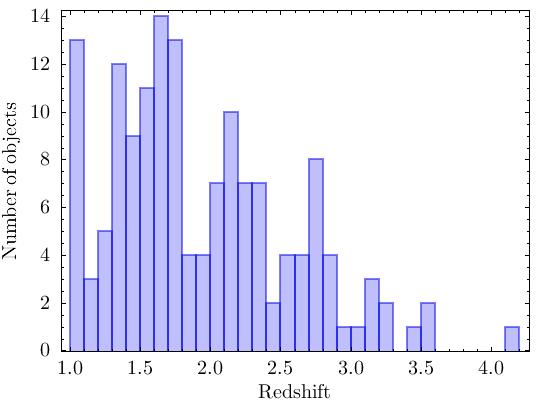}
\includegraphics[width=.48\textwidth]{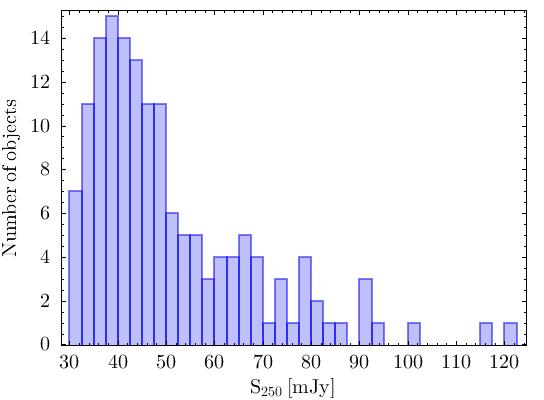}
\\
\includegraphics[width=.48\textwidth]{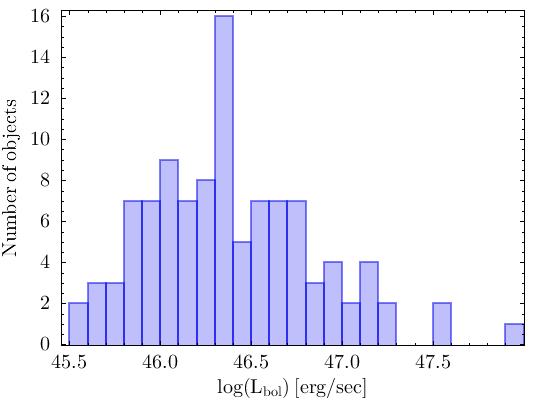}
\includegraphics[width=.48\textwidth]{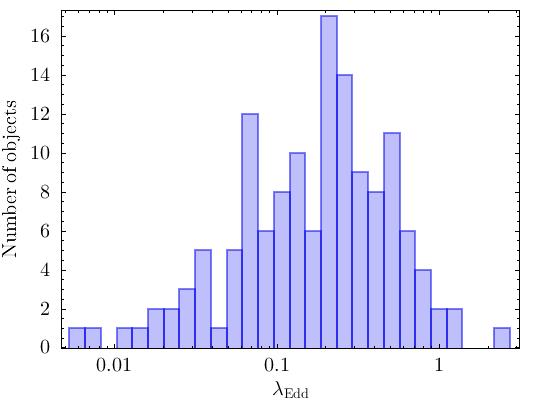}
\caption{Properties of the sample of 152 quasars. From left to right and top to bottom: spectroscopic redshift; {\it Herschel}/SPIRE 250 $\mu$m flux (S$_{250}$); quasar bolometric luminosity (L$_{bol}$) and Eddington ratio ($\lambda_{\rm Edd}$) from the SDSS DR16 catalogue.}
\label{fig:properties}
\end{figure*}

Included in this sample are High ionisation BAL (HiBAL) quasars, which exhibit distinct spectral signatures. HiBAL quasars are characterised by strong, broad absorption features in CIV at 1549 \AA, indicative of the presence of high-velocity outflows of ionised gas surrounding the quasar. MgII absorption at 2798 \AA \, is associated with low-ionisation BAL quasars (LoBAL) and is typically weaker and narrower compared to the CIV absorption.

For the SDSS DR16 quasar catalogue, an automatic procedure was put in place to identify HiBAL quasars based on the values of the Balnicity and Intrinsic Absorption indices (BI and AI, respectively), derived mainly from the CIV emission line, as described in \citep{lyke20}. Following this procedure, 17 of the quasars in our sample have BAL\_PROB=1.0 (unambiguous BAL identifications) and another 17 have BAL\_PROB=0.9 based on their AI, that are also considered as almost unambiguous BAL quasars. The BI of these quasars is zero where the trough is under 2000 km s$^{-1}$ wide and/or the trough extends closer to the line centre than 3000 km s$^{-1}$. 
We visually inspected the spectra of these 34 BAL quasars. From the 17 objects with BAL\_PROB=1.0, 15 show CIV absorption as expected, while two do not present any CIV or Si IV absorption but, instead, they show strong Ly${\alpha}$ absorption. Out of the 17 objects with BAL\_PROB=0.9, 16 present unambiguous BAL features. The remaining one is not considered hereafter. Note that only one object among the 33 shows an absorption in MgII (SDSS\_J014905.28-011404.9), and could therefore be classified as a LoBAL quasar.

\section{ALMA data}
\label{sec:almadata}

The ALMA data presented in this work were taken under Cycle 9 project 2022.1.00029.S (PI Hatziminaoglou) in Band 7. The sample was divided by the algorithm in the ALMA Observing Tool in three Scheduling Blocks (SBs), with 30, 59 and 63 objects, respectively. All were observed with the default Band~7 continuum spectral setup centred at 343.5\,GHz sky frequency (rest-frame 0.69--1.7\,THz, i.e., 435--176\,$\mu$m), with the lowest possible spectral resolution ($\sim$28\,km/s), in Time Division Mode (TDM). 

One SB was observed on October 24, 2022, the other two on October 26, 2022, in configuration C2 with 42--43 usable antennas covering baselines between 14~m and 368~m, yielding a 0.8\arcsec \, spatial resolution. The precipitable water vapour (PWV) was 0.3--0.5\,mm, while the phase-rms was $9-24$\,deg measured inter-scan on phase calibrator. Further Water Vapour Radiometer (WVR; 1\,sec scales) solutions were adopted allowing median factor improvements of 1.2--2.6, resulting in overall phase-rms values of 5--11\,deg. The expected signal decorrelation in such conditions is thus minimal ($\lesssim 2\,\%$), well within ALMA's absolute flux uncertainty of 10\% in Band~7. As a result, no self-calibration was pursued in the sample, especially because it would only be possible in a small fraction of the sample, given the fluxes of the sources (see Sect. \ref{sec:sourceident}). Each science target was observed for 48.2 seconds. Sources J2258-2758 and J0238+1636 were adopted as Bandpass calibrators, while J0108+0135 and J0217+0144 were used as Phase calibrators. All data were reduced with the ALMA pipeline, version 2022.2.0.64, with CASA version 6.4.1.12, with robust 0.5.

\section{The properties of the ALMA 870 micron sources}
\label{sec:counterparts}

\subsection{Identification of submm sources}
\label{sec:sourceident}

We automated the source extraction procedure with a custom-made python script, that includes the identification of secondary counterparts, submm flux measurement and fitting on the ALMA Band 7 continuum maps. The steps in the script are as follows:
\begin{enumerate}
    \item Computation of noise levels, i.e. Median Absolute Deviation (MAD) and standard deviation (RMS) using only pixels within $\pm5 \times$MAD;
    \item Identification of peak-fluxes above 5$\sigma$; 
    \item \label{ext:boxphot} Extraction of the flux density within a box-shaped aperture centred on the peak pixel and with a width twice the beam major axis;
    \item For each detected source, finding of the peak flux and signal-to-noise ratio, computation of the light-weighted centroid of the intensities above 3$\sigma$, and measurement of the distance and position angle from the phase centre;
    \item Use of the CASA {\sc imfit} task to fit the emission within a elliptical aperture twice the beam size (but with the same major to minor axis ratio and position angle), centred on the light-weighted centroid;
    \item For the quasars with no peak-fluxes above 5$\sigma$, photometry extraction at the phase centre was forced as described in point \ref{ext:boxphot}, with an attempt to run {\sc imfit} only when the peak-flux was above 3$\sigma$.
\end{enumerate}

\begin{table*}[]
\centering
\caption{First five entries of the ALMA Band 7 photometric catalogue in the fields of the 152 quasars.}
\setlength{\tabcolsep}{6pt}
\begin{tabular}{lccrrrcccc}
\hline
SDSS ID & RA$_{\rm wgt}$ & Dec$_{\rm wgt}$ & D2P & SNR 
& $I_{\nu}^{\rm ap}$ & S$_{\nu}^{\rm ap}$ & $I_{\nu}^{\rm imfit}$ & S$_{\nu}^{\rm imfit}$ \\
& [h:m:s] & [d:m:s] & [\arcsec] & peak & [mJy/beam] & [mJy] & [mJy/beam] & [mJy] \\
\hline
  J005624.62--000438.5 & 00:56:24.62 & -00:04:38.5 & 0.11 & 12 & 1.90$\pm$0.16 & 1.94$\pm$0.30 & 1.98$\pm$0.62 &  2.11$\pm$0.30\\
  J005642.28+000104.7 & 00:56:42.29 & +00:01:04.7 & 0.05 & 27 & 4.51$\pm$0.17 & 4.85$\pm$0.31  & 4.50$\pm$0.17 & 5.20$\pm$0.33\\
    & 00:56:41.90 & +00:01:03.1 & 5.9 & 13 & 2.89$\pm$0.23 & 3.16$\pm$0.31 & 2.95$\pm$0.23 & 3.40$\pm$0.44\\
  J005713.03+005725.5 & 00:57:13.04 & +00:57:25.5 & 0.08 & 16 & 2.88$\pm$0.18 & 3.54$\pm$0.33 & 2.82$\pm$0.19 & 4.01$\pm$0.42\\
  J005743.78--001158.9 & 00:57:43.78 & -00:11:59.0 & 0.05 & 23 & 3.73$\pm$0.16 & 3.96$\pm$0.30 & 3.63$\pm$0.17 & 4.28$\pm$0.32\\
\hline
\end{tabular}
\tablefoot{The columns are as follows: SDSS Name, coordinates (RA$_{\rm wgt}$, Dec$_{\rm wgt}$) of the light centroid position, the distance to the phase centre (D2P), the signal-to-noise (SNR) at the peak, the peak intensity and integrated flux ($I_{\nu}$ and S$_{\nu}$, respectively), for aperture photometry (ap) and from {\sc imfit}, and $\sigma$ of the extractions.}
\label{tab:contphot}
\end{table*}

The above procedure resulted in the extraction of 209 Band 7 sources in the 152 quasar fields, described in the next section. For 135 out of the 152 quasars (89\%)
the counterparts were centred on the optical coordinates with peak fluxes above 5$\sigma$. For the 17 quasars not detected above 5$\sigma$, forced photometry was performed at the location of the quasar, with 7 of those at or above 3$\sigma$. We consider the remaining 10 objects with peak fluxes below $3\sigma$ as non-detections in the submm and are not included in what is discussed below. 

\begin{figure*}[]
\center
\includegraphics[width=0.48\textwidth]{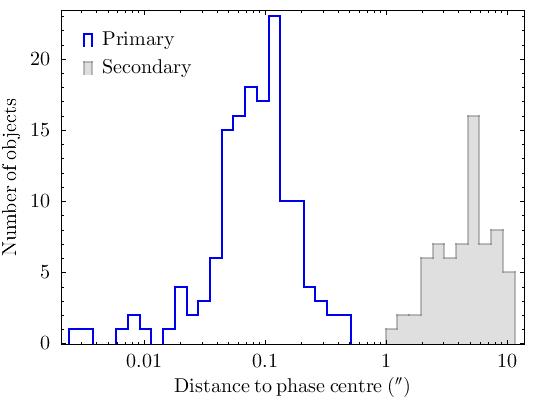}
\includegraphics[width=0.48\textwidth]{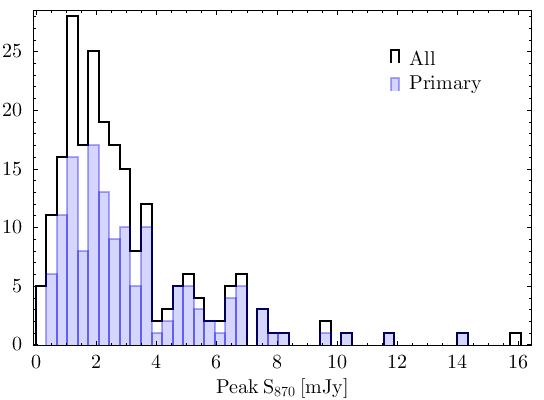} 
\caption{{\it Left}: Histogram of the distance to the phase centre for the primary (open histogram) and secondary (grey histogram) counterparts. {\it Right}: Distribution of the Band 7 (870 \mum) peak fluxes of the 209 ALMA sources in the 152 quasar fields (black histogram). Primary counterparts, i.e. counterparts extracted at the phase centre (i.e. at the optical coordinates of the quasars) are shown in a shaded histogram.}
\label{fig:histos}
\end{figure*}

\begin{figure*}
\centering
\begin{tabular}{lll}
\setlength{\tabcolsep}{0pt}
\includegraphics[width=0.32\textwidth]{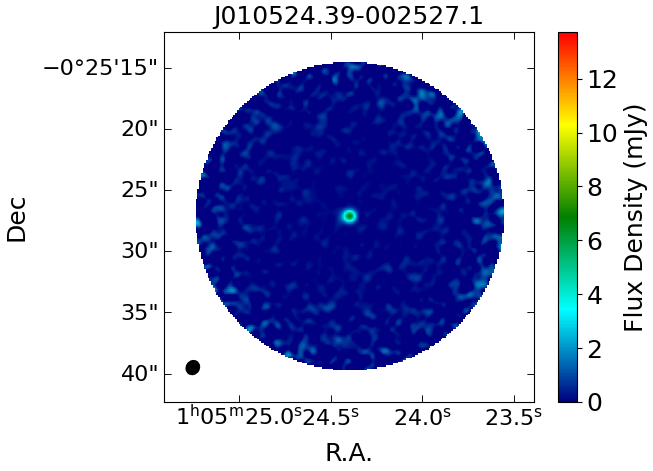} & 
\includegraphics[width=0.32\textwidth]{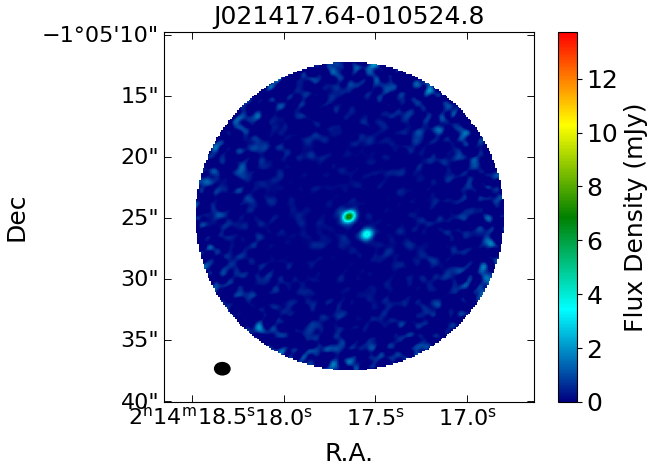} &
\includegraphics[width=0.32\textwidth]{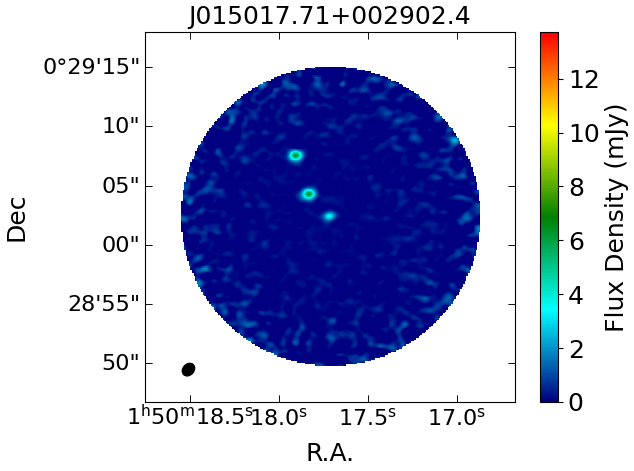} \\
\includegraphics[width=0.32\textwidth]{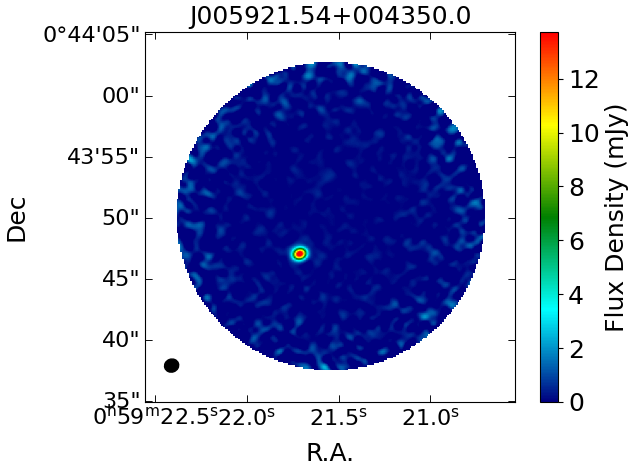} & 
\includegraphics[width=0.32\textwidth]{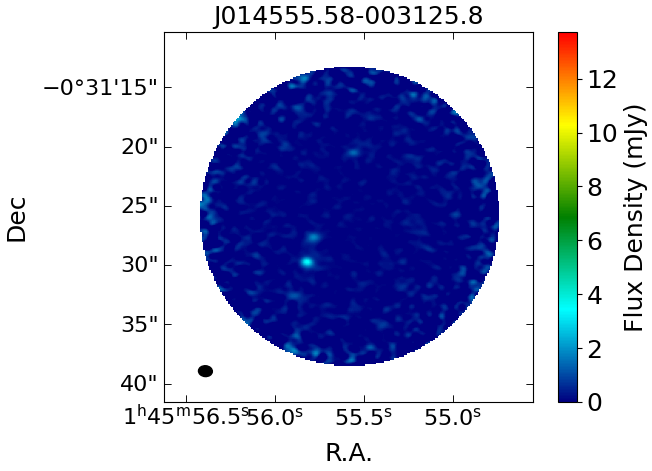} &
\includegraphics[width=0.32\textwidth]{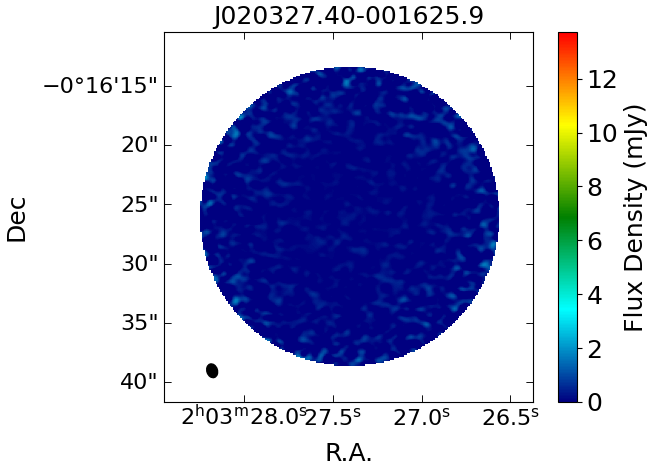} \\
\end{tabular}
    \caption{Example ALMA Band 7 images. From top to bottom and left to right: J010524.39-002527.1: Single 870 \mums counterpart centred on the SDSS coordinates; J021417.64-010524.8: ALMA 870 \mums counterpart centred at the location of the quasar, with secondary counterpart; J015017.71+002902.4: ALMA 870 \mums counterpart centred at the location of the quasar, with two secondary, brighter counterparts; J005921.54+004350.0: Single 870 \mums counterpart, not associated with the SDSS quasar; J014555.58-003125.8: Three 870 \mums counterpart, none of which associated with the quasar; and J020327.40-001625.9: no detection. The ALMA beam is shown as a black ellipse at the bottom left corner of each image.
    }
\label{fig:cutouts}
\end{figure*}

Out of 152 quasar fields, 94 (62\%) have a single submm counterpart, with two of these located more than 1\arcsec \, away from the central quasar's coordinates and are therefore not related to the primary source. 
In 42 fields (28\%), two submm counterparts are detected, while nine fields (6\%) host three counterparts. One field contains four submm sources. In one of the fields with two counterparts and in one with three, the central quasar was not detected at 870 \mum.
Secondary counterparts are located between 1\arcsec \, and $\sim$10\arcsec \, from the quasar's phase centre (Fig. \ref{fig:histos}, left panel), corresponding to projected physical distances between $\sim$10 and $\sim$90 kpc. The right panel of Fig. \ref{fig:histos} displays the 870 \mums flux distribution for the 209 submm sources (open black histogram). The shaded histogram includes the primary sources only, i.e. the 142 counterparts coinciding within $<0.5$\arcsec \, from the optical coordinates of the quasars. Table \ref{tab:contphot} shows the first five entries of the list of submm detections. The columns of the table are described in the table caption. The full table is available at the CDS. Figure \ref{fig:cutouts} illustrates examples of submm detections in six fields.

The 10 quasars (7\% of the sample) with no primary submm counterpart are noteworthy, as they may suggest issues with the FIR data or the association of FIR counterparts to the optical sources. Of those 10 cases, one has no 500 \mums detection and five of them have 500 \mums fluxes below 11.5 mJy (the median of the sample is 22.5 mJy). The remaining four non-detections are in fields with secondary submm sources, suggesting that the FIR emission likely came from unrelated sources rather than the quasar host.

\subsection{Multiplicities as a function of FIR properties, redshift and ``balnicity''}
\label{sec:fluxzbal}

Studying the multiplicity of 250 \mums {\it Herschel} sources in the COSMOS field and using Bayesian interference methods, \cite{scudder16} reported that, in the presence of more than one counterparts, for sources with 250 \mums fluxes above 45 mJy, the brightest and second brightest components are assigned comparable fluxes, the sum of which does not reach the total 250 \mums flux of the SPIRE source. For the fainter objects with 250 \mums fluxes below 45 mJy, the majority of the 250 \mums flux comes from a single bright component. The second brightest component, when present, is typically a lot fainter.

\begin{figure}
    \centering
    \includegraphics[width=0.5\textwidth]{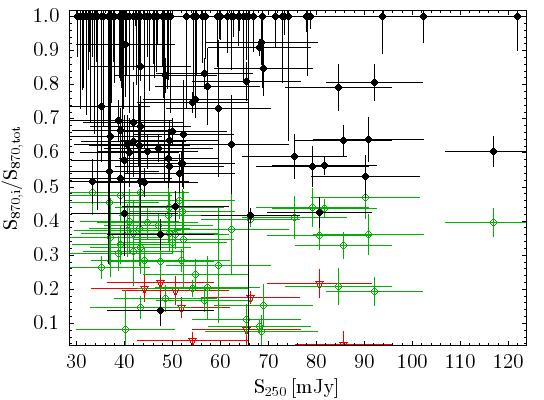}
    \caption{Contribution of the brightest (black filled circles), second  and third brightest (green open circles and red open triangles, respectively) counterparts to the total 870 \mums flux on each of the ALMA Band 7 maps, as a function of the 250 \mums originally associated to each SDSS quasar.}
    \label{fig:frac870vss250}
\end{figure}

The above results are not in agreement with our findings. Looking into the contribution of the multiple submm counterparts to the total 870 flux on each of the maps, we find indeed that the majority of objects (48 of the 74 or 64.9\% $\pm$ 9.4\%) with S$_{250}<45$ mJy have a single submm counterpart. In the presence of a secondary submm counterpart, the brightest 870 \mums source accounts for 60\% of the 870 \mums flux, on average, within the SPIRE beam, with the secondary source accounting for 42\% $\pm$ 4\% of the total submm flux. The picture is very similar for the bright SPIRE sources with S$_{250}>45$ mJy. We find that 44 out of the 78 (56.4\% $\pm$ 8.5\%) have single submm counterparts. Once again, when a secondary counterpart is present, the primary counterpart accounts, on average, for 60\% of the submm flux, with the secondary contributing by 37\% $\pm$ 10\% to the total submm flux. Of the 78 sources with S$_{250}>45$ mJy, only 8 have a tertiary counterpart, the contribution of which to the total 870 flux does not exceed 20\%. In summary, we do not see any significant change in the number of submm counterparts as a function of 250 \mums flux, while, in the presence of multiple counterparts, the contribution of the primary counterpart to the total submm flux is of the order of 60\%, with the secondary accounting for the remaining flux with a few exceptions, that require a third counterpart that contributes up to 20\% to the total flux. The above are illustrated in Fig. \ref{fig:frac870vss250}, that is equivalent to Fig. 4 in \cite{scudder16}.

As multiplicities might indicate over-dense environments around the central quasar, we also checked them as a function of redshift. Table \ref{tab:zmulti} shows the number of non-detections (discussed above) and of quasars with more than one submm counterparts in their respective fields. We note an increase by a factor of $\sim2$ or more from redshift 1 to 2.5 in the multiplicity incidence. Last but not least, no dependence of the multiplicity is found with ${\rm \lambda}_{Edd}$. The implications of the above findings are discussed in Sect. \ref{sec:discuss}. 

\begin{table}
    \centering
    \caption{Non-detections and multiple components per redshift bin, with Poissonian errors.}
    \setlength{\tabcolsep}{4pt}
    \begin{tabular}{cccc}
    \hline
        Redshift & \# Objects & Non-detections & Multiple \\
        \hline
        1 - 1.5 & 42 & 5 & \, 9 (21.4\% $\pm$ \,7.1\%)\\
        1.5 - 2 & 46 & 1 & 12 (26.1\% $\pm$ \,7.5\%)\\
        2 - 2.5 & 33 & 2 & 17 (51.5\% $\pm$ 12.5\%)\\
        $>$ 2.5 & 31 & 2 & 16 (51.6\% $\pm$ 12.9\%)\\
        \hline
    \end{tabular}
    \label{tab:zmulti}
\end{table}

The sample of 152 quasars includes 33 visually confirmed HiBAL quasars, 17 with BAL\_PROB=1.0 and 16 with BAL\_PROB=0.9 (see Sect. \ref{sec:sdsssample}). 
From the objects with BAL\_PROB=1.0, seven (41\% $\pm$ 16\%) have at least one secondary counterpart, among which the two with strong Ly${\alpha}$ absorption. Among the 16 objects with BAL\_PROB=0.9, six (38\% $\pm$ 15\%) have at least one secondary counterpart. The only LoBAL quasar in the sample (SDSS\_J014905.28-011404.9) does not have a secondary submm counterpart. The incidence of multiplicities among the BAL quasars does not differ from that observed in the full sample, demonstrating that previous claims on a dependency with ``balnicity'' were driven by low number statistics. 

\section{Serendipitous line detections}
\label{sec:lines}

Given the broad redshift range and the size and nature of the sample and despite the very coarse spectral resolution, it is expected that some of the main CO transition will be present in the spectra of some of the targets, redshifted into the survey spectral coverage. In particular, the CO transition from CO(6-5) to CO(15-14) fall within the frequency range covered by our sample. For the redshifts of the objects in the sample, with a redshift uncertainty of 0.002 (typical for SDSS quasars, as derived from Principal Component Analysis methods; \citealt{lyke20}), and considering the Band 7 continuum spectral setup (which consists of four spectral windows, each 1.875 GHz wide, centred at 336.5 GHz, 338.5 GHz, 348.5 GHz, and 350.5 GHz), a total of 31 CO transitions fall within the spectral range covered by the ALMA observations. These include three CO(6-5), five CO(7-6), eight CO(8-7), five CO(9-8), and ten CO(10-9) and higher-J transitions.

\begin{figure*}
    \centering
    \begin{minipage}[t]{0.6\textwidth}
        \centering
        \vspace{0pt}
        \includegraphics[width=\textwidth]{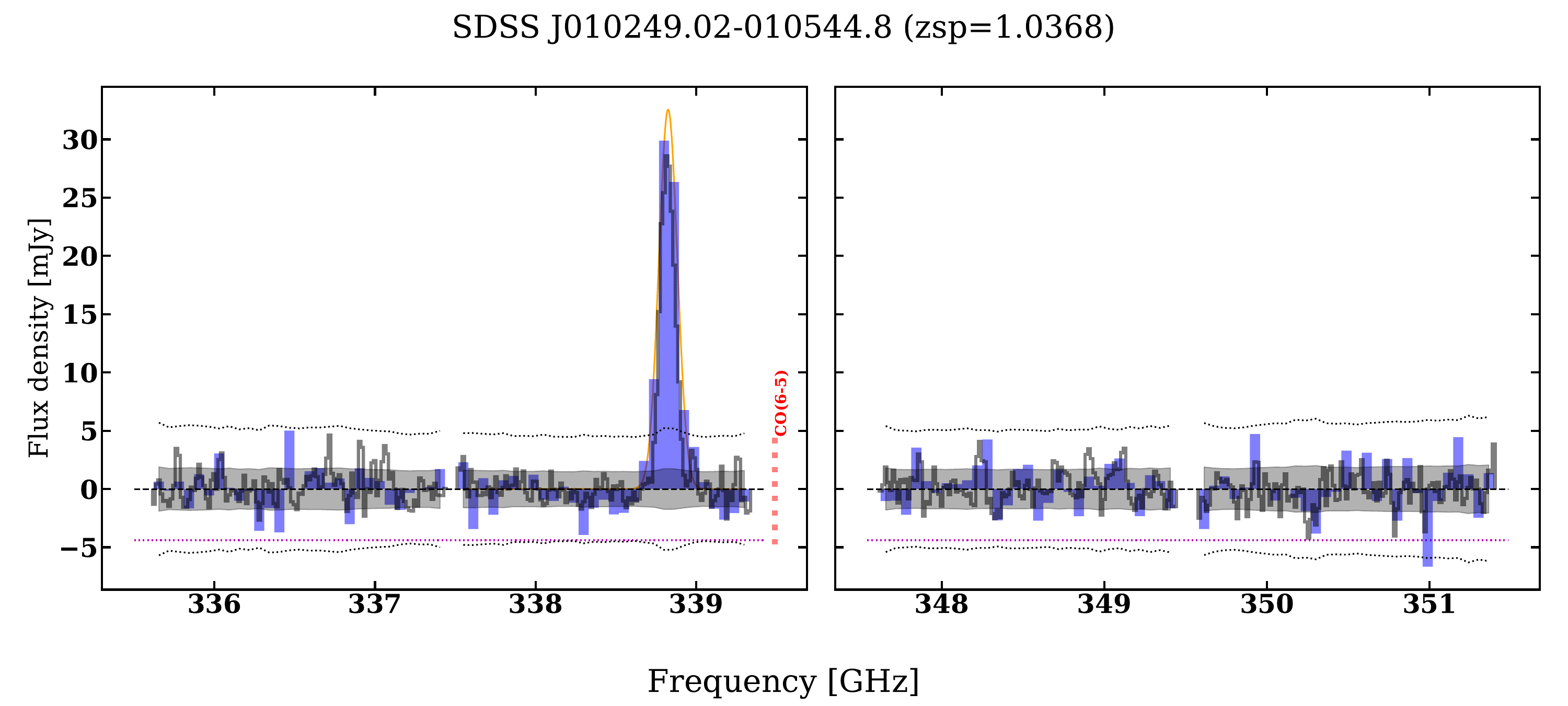}
    \end{minipage}
    \begin{minipage}[t]{0.32\textwidth}
        \centering
        \vspace{0pt}
        \includegraphics[width=\textwidth]{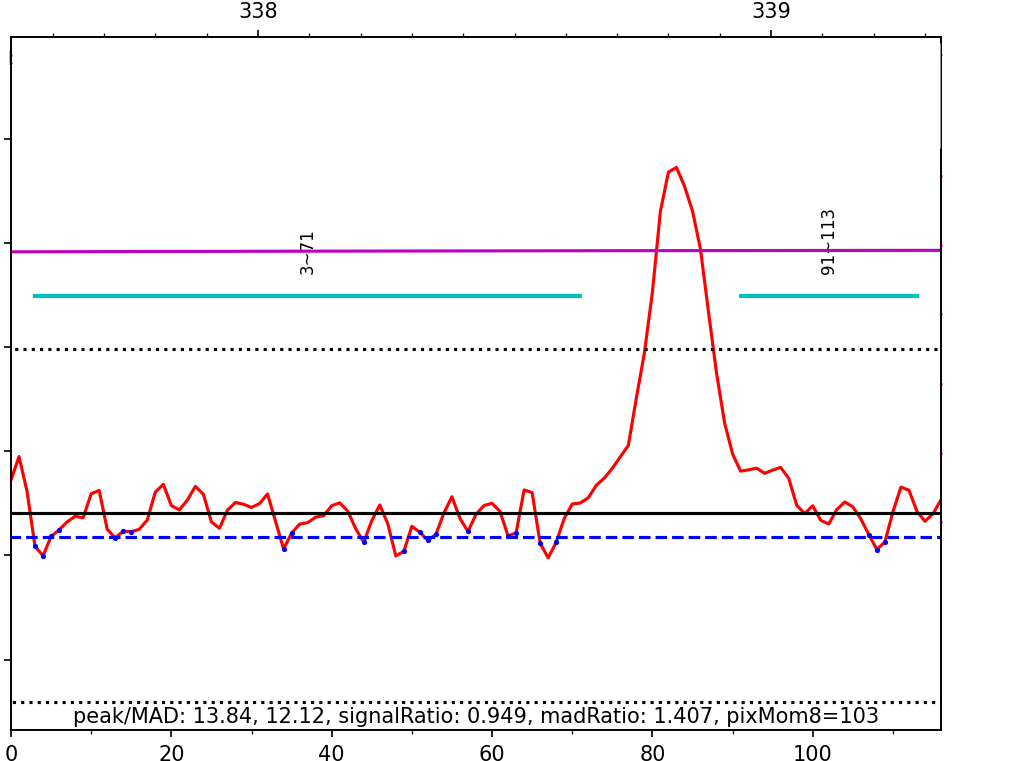}
    \end{minipage}
    \\
    \begin{minipage}[t]{0.6\textwidth}
        \centering
        \vspace{0pt}
        \includegraphics[width=\textwidth]{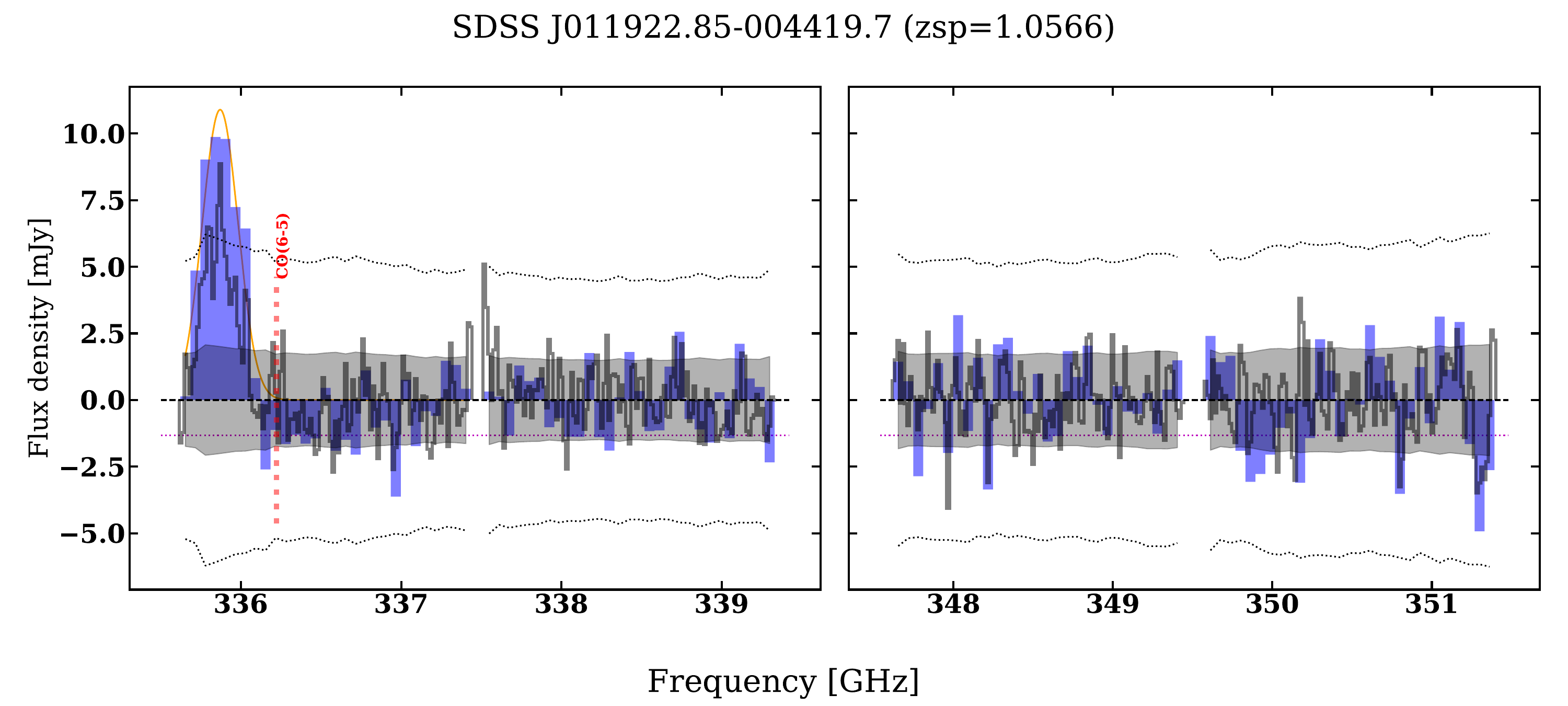}
    \end{minipage}
    \begin{minipage}[t]{0.32\textwidth}
        \centering
        \vspace{0pt}
        \includegraphics[width=\textwidth]{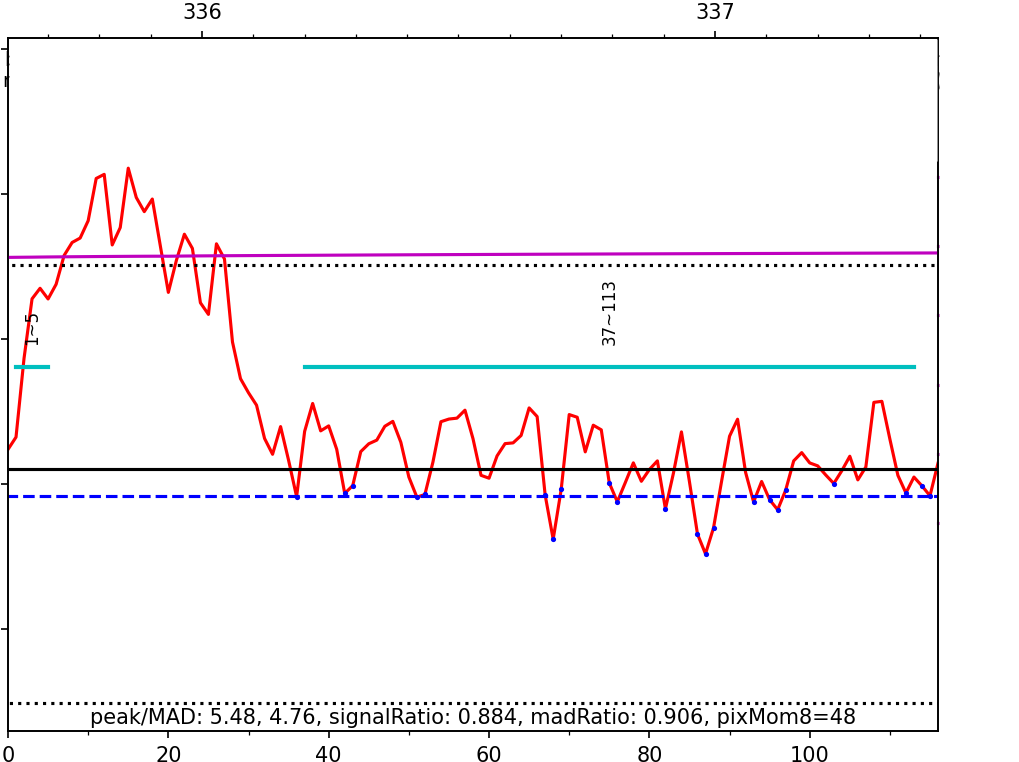}
    \end{minipage}
    \caption{The Band 7 spectra of the two quasars with unambiguous CO(6-5) transition detections. {\it Left and middle columns:} The filled blue histograms show the extracted spectra within the extraction aperture smoothed to a spectral resolution of $\sim$111\,km s$^{-1}$, while the solid black histogram shows the raw-resolution spectra of the peak flux within the aperture (units of Jy/beam). The greyed out regions are the per-channel $\pm1\sigma$, while the dotted lines are the per-channel $\pm3\sigma$ level. The magenta line shows the continuum that was subtracted from the spectrum to assess the significance of absorption detections (there were none). The red dashed lines indicate the expected location of the CO transition shifted to the optical (SDSS) redshift of each object. The orange line shows the single-Gaussian fit to the line.
    {\it Right column:} ALMA pipeline continuum fitting output from the spectral window where the lines were identified. The binned spectrum in red; the region in which the continuum has been extracted is shown in cyan; the level of the continuum is shown in black. The x-axis at the bottom indicates the channels, the one at the top the frequency (in GHz). These plots have been extracted directly from the logs of the pipeline runs ("weblogs"). 
}
\label{fig:linesearch1}
\end{figure*}

ALMA spectra were extracted for all 209 detections  
and were visually inspected. The extraction was performed within an aperture with the size and position angle of the synthesised beam. 
A detection was considered reliable if the peak of the binned spectrum exceeded the 3$\sigma$ per channel level and was supported by the existence of a nearby transition within a velocity shift compatible with the error on the redshift (see also the caption of Fig. \ref{fig:linesearch1}).
A visual inspection was also performed on all the plots produced by the ALMA pipeline runs at the step of continuum finding, available in the pipeline logs. 

No transition higher than CO(7-6) were identified in the low-resolution ALMA spectra. Two of the three expected CO(6-5) lines (rest frequency of 691.473 GHz) are clearly visible in the spectra of the objects (SDSS\_J010249.02-010544.8 and SDSS\_J011922.85-004419.7). The third one (SDSS\_J013109.99+010612.8) has no sign of emission on its spectrum and it is, in fact, undetected at 870 \mums with no other submm sources detected in its vicinity. The quasar has an estimated SFR (derived by \citealt{pitchford16}) of $\sim$520 M$_{\odot} yr^{-1}$ and is accreting at a $\lambda_{\rm edd}$ of 0.03, both at the very low end of the parameter distribution in the sample. Figure \ref{fig:linesearch1} shows the extracted spectra split by baseband in the left and middle panels. The right panel shows the ALMA pipeline continuum fitting output from the spectral window in which the identified transition is located. 

\begin{figure*}
    \centering
    \begin{minipage}[t]{0.6\textwidth}
        \centering
        \vspace{0pt}
        \includegraphics[width=\textwidth]{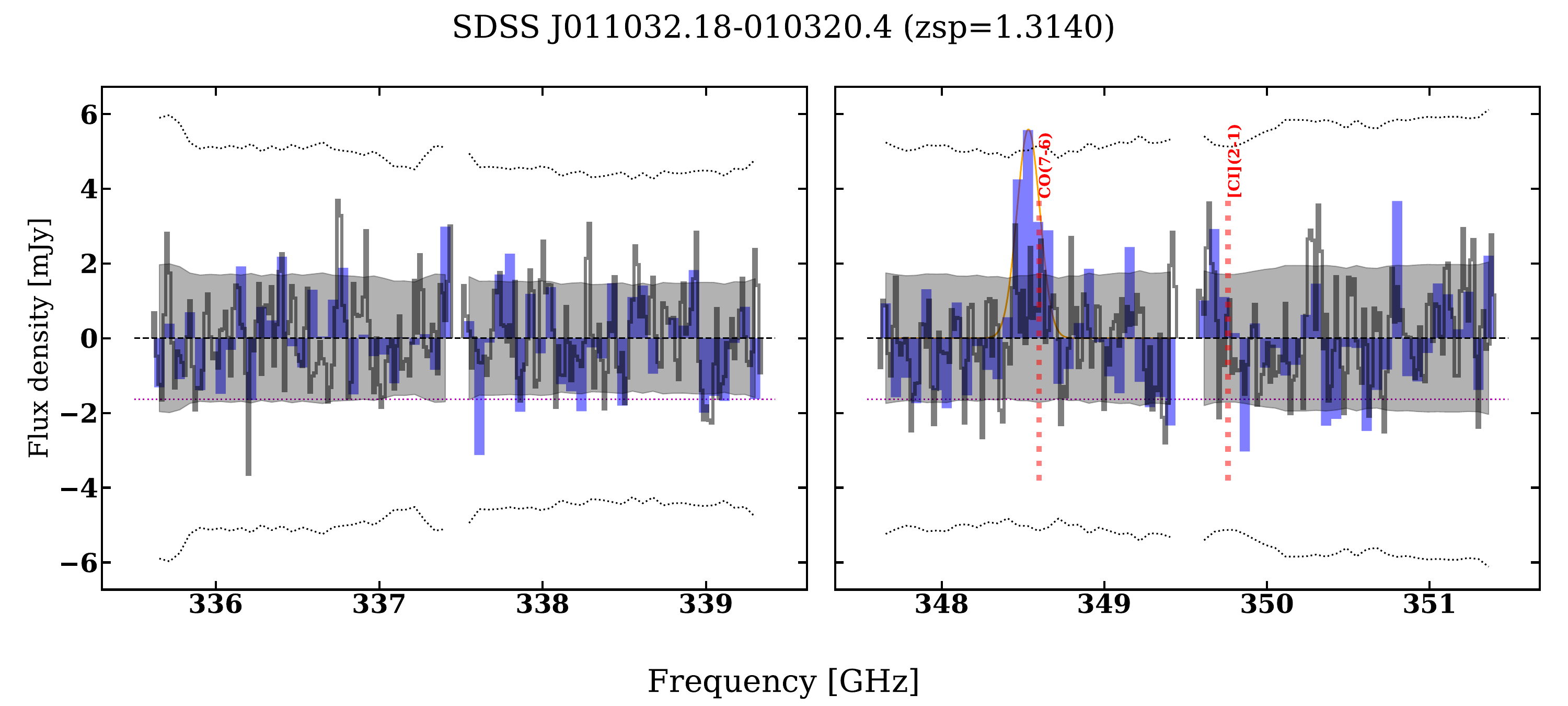}
    \end{minipage}
    \begin{minipage}[t]{0.32\textwidth}
        \centering
        \vspace{0pt}
        \includegraphics[width=\textwidth]{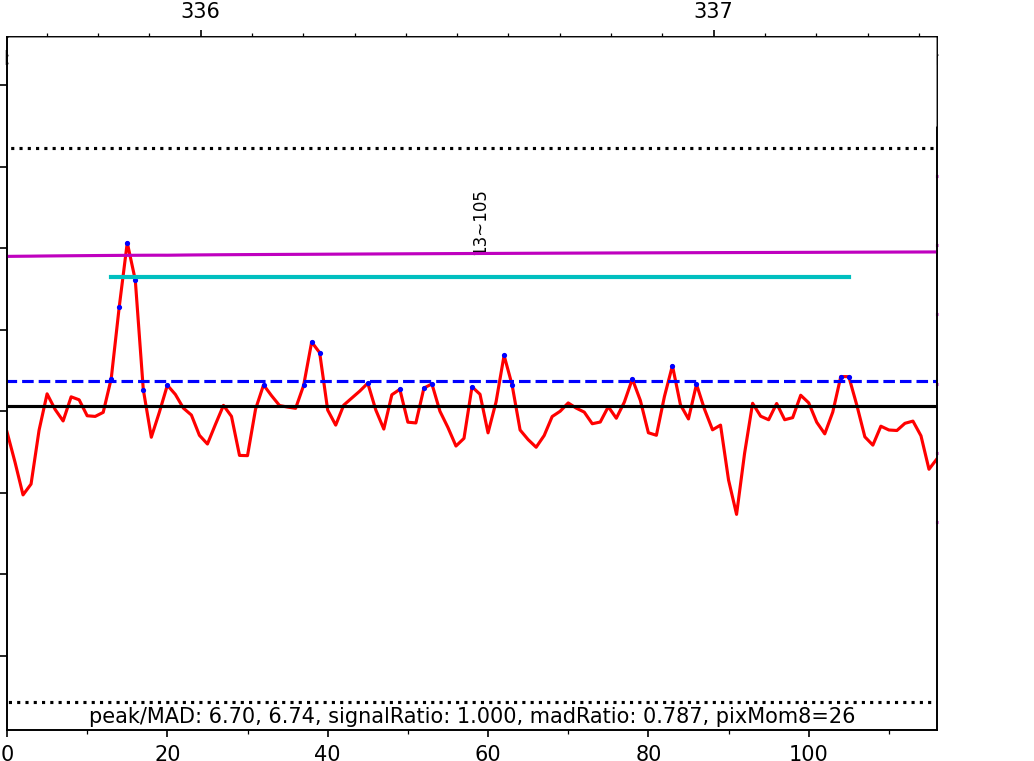}
    \end{minipage}
    \\
    \begin{minipage}[t]{0.6\textwidth}
        \centering
        \vspace{0pt}
        \includegraphics[width=\textwidth]{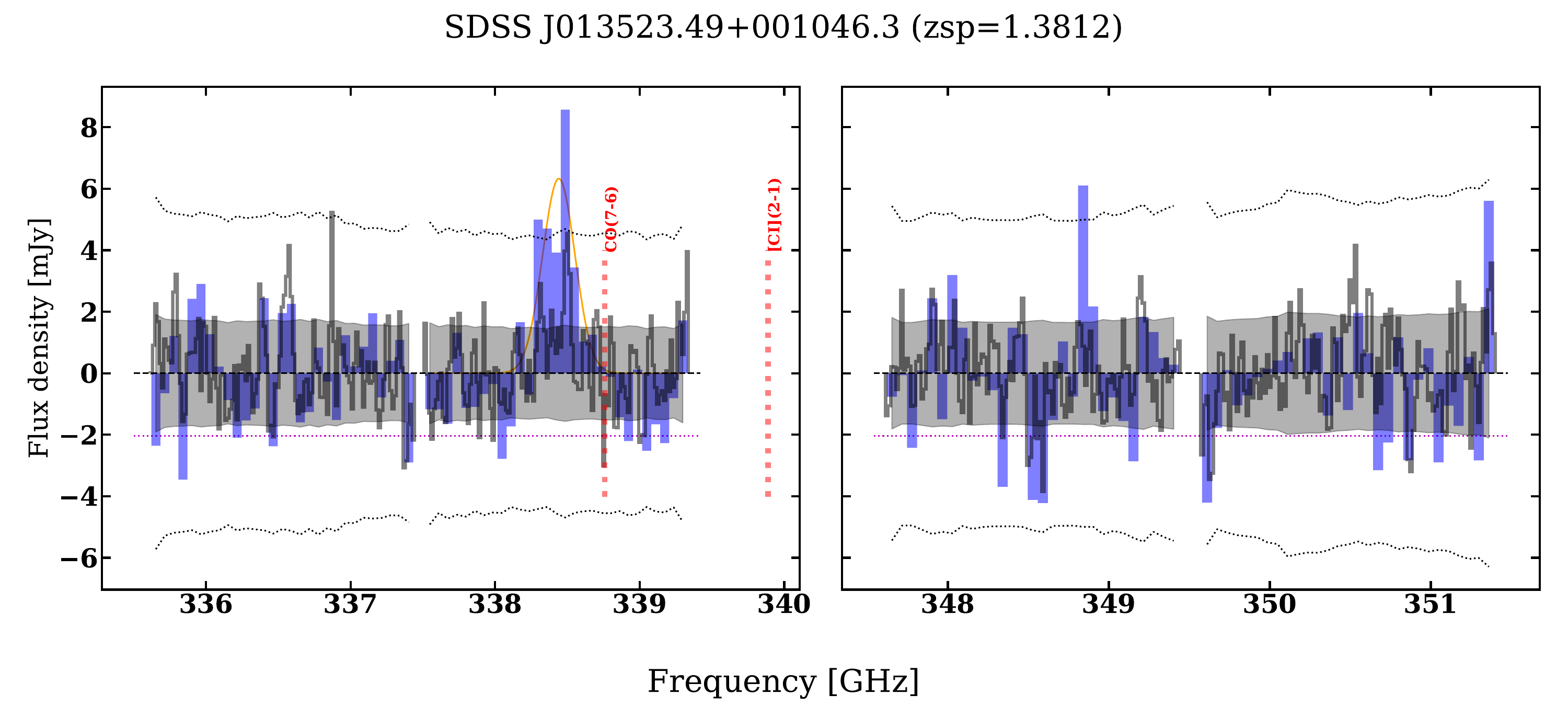}
    \end{minipage}
    \begin{minipage}[t]{0.32\textwidth}
        \centering
        \vspace{0pt}
        \includegraphics[width=\textwidth]{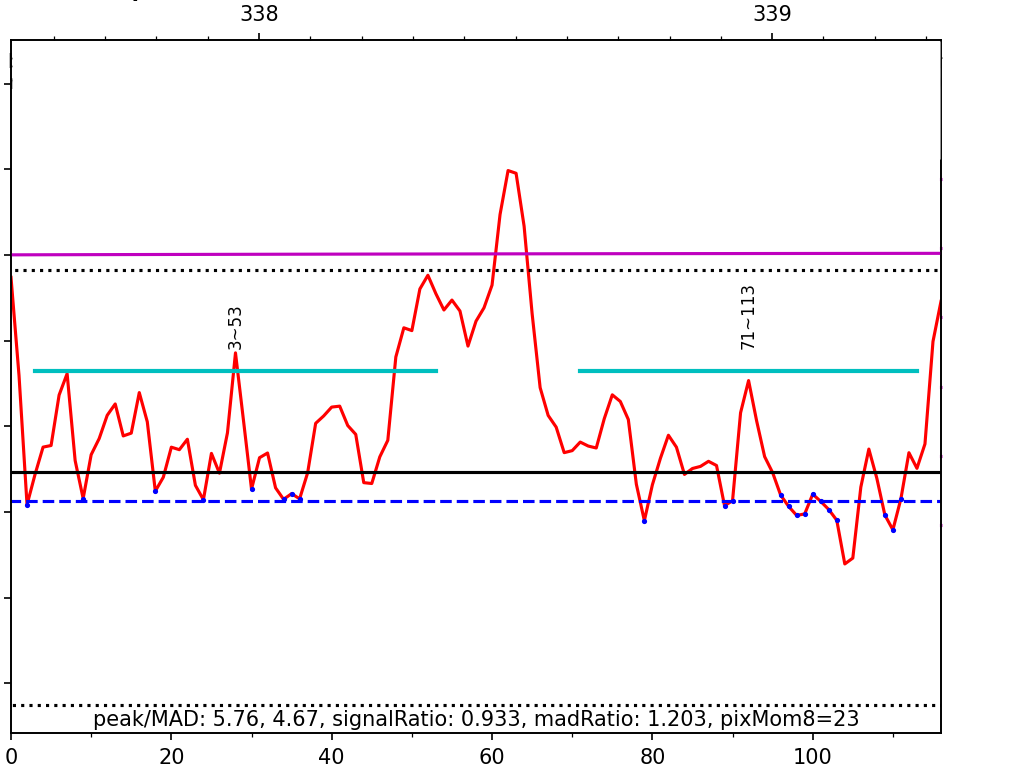}
    \end{minipage}
    \\
    \begin{minipage}[t]{0.6\textwidth}
        \centering
        \vspace{0pt}
        \includegraphics[width=\textwidth]{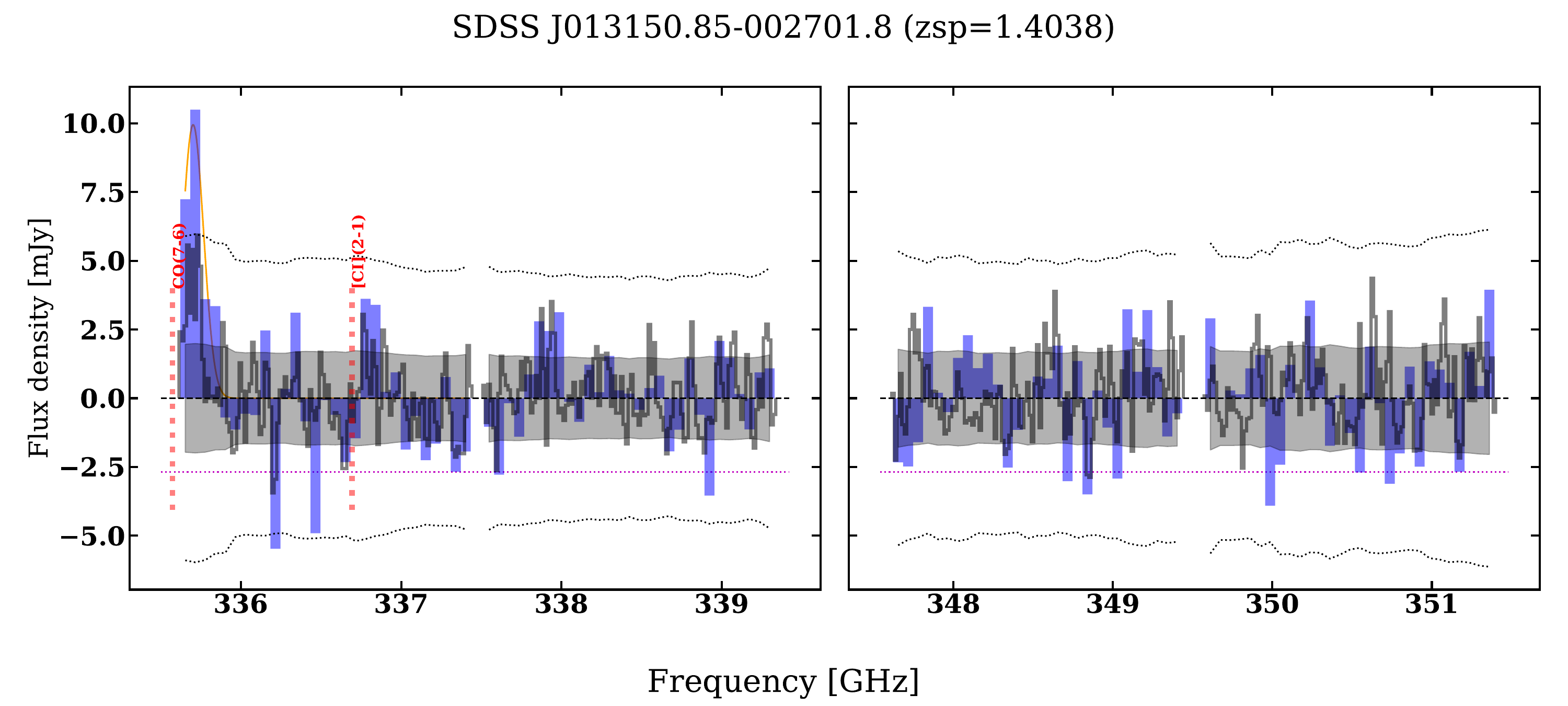}
    \end{minipage}
    \begin{minipage}[t]{0.32\textwidth}
        \centering
        \vspace{0pt}
        \includegraphics[width=\textwidth]{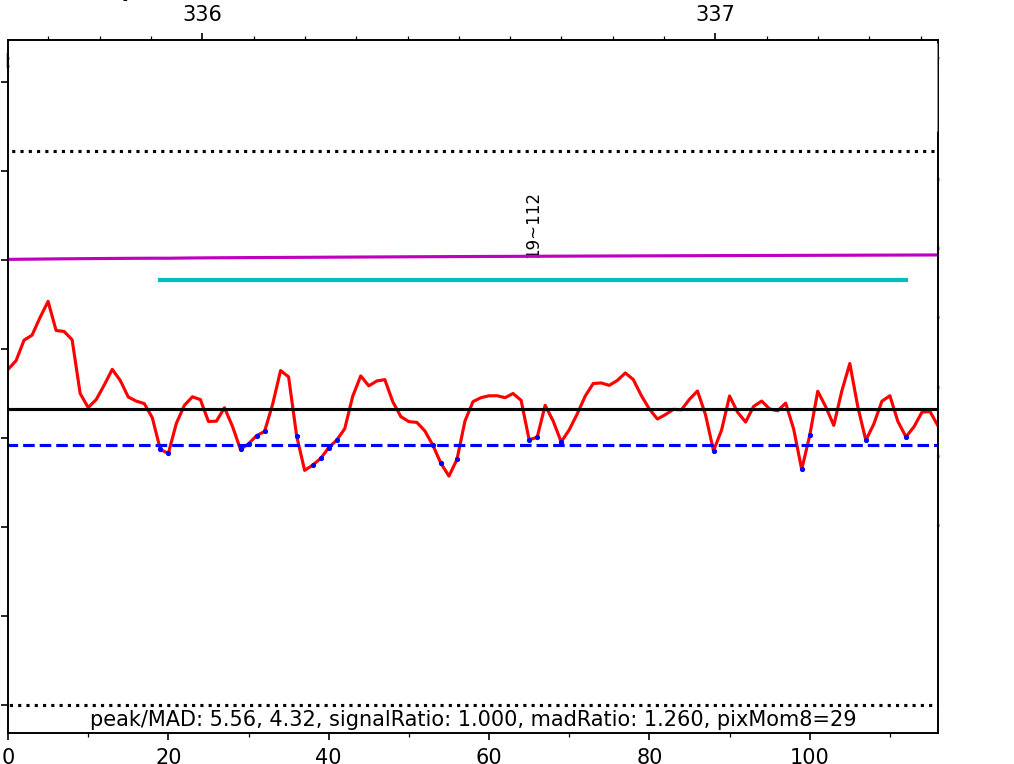}
    \end{minipage}
\caption{Same as Fig. \ref{fig:linesearch1} but for CO(7-6) transitions. The expected location of the nearby CI(2-1) transition is also marked on the plots, however these transitions are not detected at a significant level. 
}
\label{fig:linesearch2}
\end{figure*}

Out of the five expected CO(7-6) transitions (rest frequency 806.651 GHz), three are detected in the spectra (SDSS\_J013523.49+001046.3, SDSS\_J013150.85-002701.8 and SDSS\_J011032.18-010320.4). These are shown in Fig. \ref{fig:linesearch2}. The remaining two that were not detected were associated to quasars both detected at above 5$\sigma$ on the Band 7 maps, one of which with a secondary counterpart lying at 6.6\arcsec \, distance. 

The five detected CO transitions were all shifted with respect to the frequency corresponding to the optical (SDSS) redshift by between 0.1 and 0.5 GHz, equivalent to velocity shifts between 85 and 440 km s$^{-1}$. These shifts equate to redshift differences of 0.002 or below, within the uncertainties of the SDSS redshift values. The transitions, observed frequencies, significance of the detections and integrated fluxes are shown in Table \ref{tab:colines}.

\begin{table*}
    \centering
    \caption{CO line identification towards the five sources shown in Figures~\ref{fig:linesearch1} and \ref{fig:linesearch2}.}
    \begin{tabular}{lccccccc}
    \hline
        SDSS ID & $z_{\rm SDSS}$ & CO & $\nu_{obs}$ & $\Delta_{vel}$ & $\sigma_{peak}$ & FWHM$_{\rm CO}$ & S$_{{\rm CO} fit}$\\
         & & transition & [GHz] & [km s$^{-1}$] & & [km s$^{-1}$] & [Jy.km s$^{-1}$] \\
        \hline
        J010249.02-010544.8 & 1.037 & 6-5 & 339.98 & 440 & 17.2 & 84.7 $\pm$ 4.7 & 2.935 $\pm$ 0.150 \\
        J011922.85-004419.7 & 1.057 & 6-5 & 336.43 & 270 & \,\, 4.9 & 166.8 $\pm$ 23.2 & 1.934 $\pm$ 0.243 \\
        J011032.18-010320.4 & 1.314 & 7-6 & 348.71 & \,\,\,85 & \,\, 3.3 & 101.6 $\pm$ 30.0 & 0.604 $\pm$ 0.156 \\
        J013523.49+001046.3 & 1.381 & 7-6 & 339.18 & 350 & \,\, 5.5 & 165.6 $\pm$ 31.1 & 1.115 $\pm$ 0.184\\
        J013150.85-002701.8 & 1.404 & 7-6 & 335.58 & -90 & \,\, 5.3 & 97.6 $\pm$ 27.6 & 1.034 $\pm$ 0.195\\
        \hline
    \end{tabular}
    \tablefoot{The observed frequency ($\nu_{obs}$), the velocity offset of the CO line with respect to $z_{\rm SDSS}$ ($\Delta_{vel}$), the significance of the detection peak ($\sigma_{peak}$), the FWHM of the Gaussian fit (FWHM$_{\rm CO}$) and the integrated line flux from the Gaussian fit (S$_{{\rm CO} fit}$) are reported.}
    \label{tab:colines}
\end{table*}

From the three objects with expected CO(6-5) lines and the five with expected CO(7-6) lines only one has a secondary counterpart, lying at a distance of 6.6\arcsec \, from the phase centre. 
No lines were present on the spectrum of this source and no secondary counterparts showed convincing line emission in the automated search. 

\begin{figure*}
    \centering

    \begin{minipage}[t]{0.6\textwidth}
        \centering
        \vspace{0pt}
        \includegraphics[width=\textwidth]{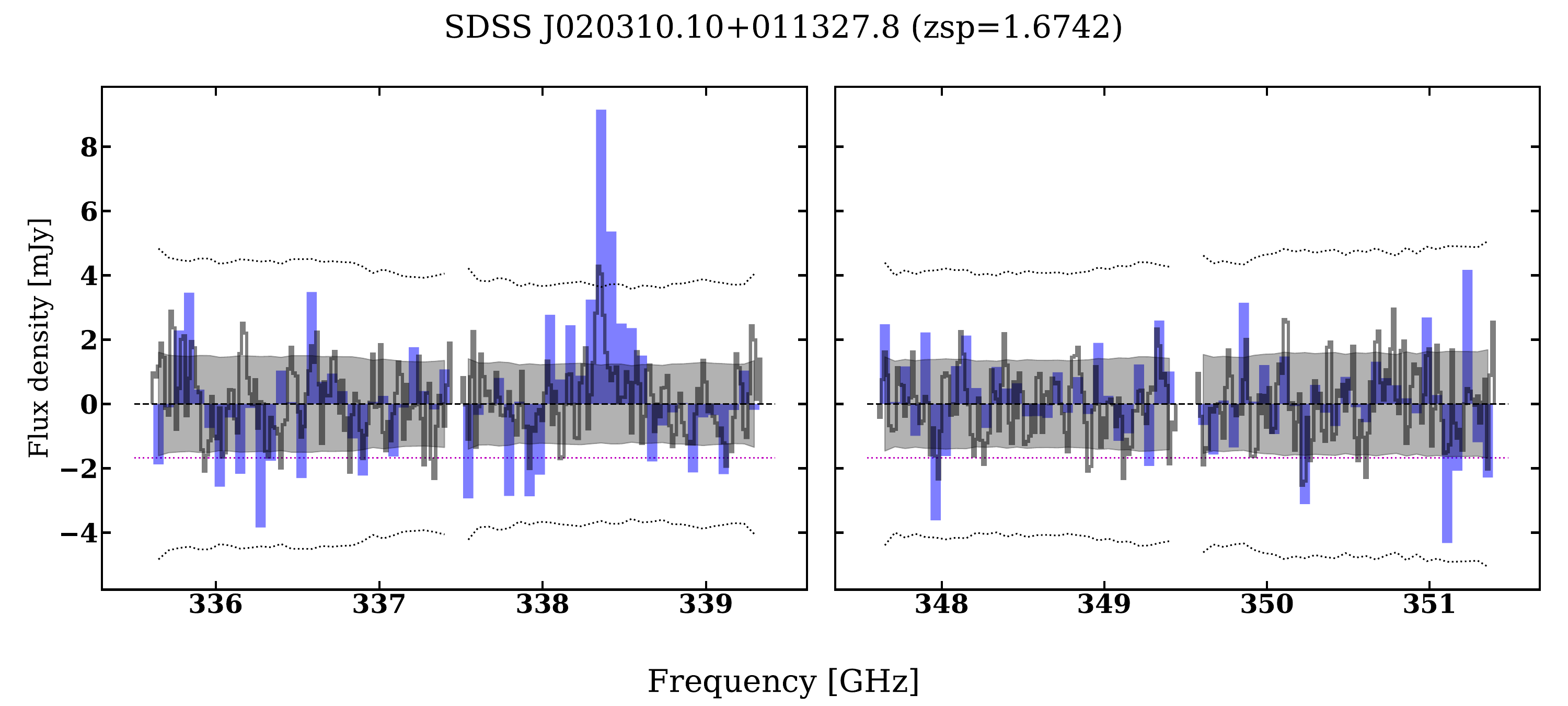}
    \end{minipage}
    \begin{minipage}[t]{0.32\textwidth}
        \centering
        \vspace{0pt}
        \includegraphics[width=\textwidth]{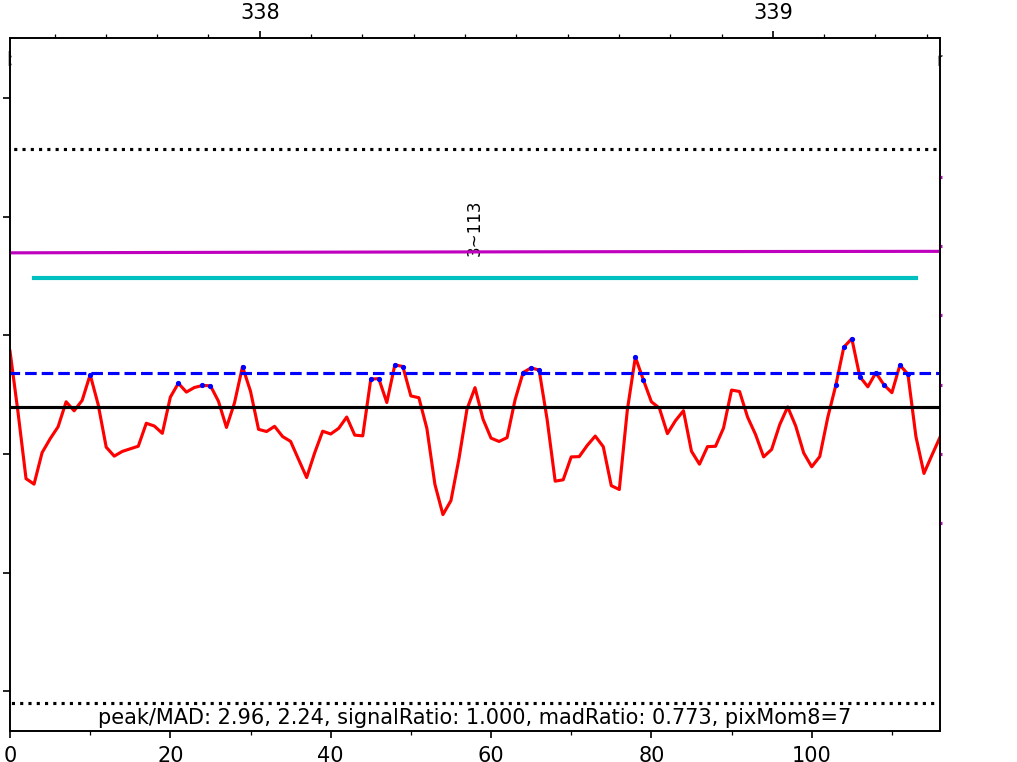}
    \end{minipage}
    \caption{H$_2$O~$9_{2, 8}- 8_{3, 5}$ or HNC~(10-9) or a mix of both in the spectrum of SDSS\_J020310.10+011327.8. The line was not picked up by the continuum finder of the ALMA pipeline nor is it visible in the weblog graph (right panel). No Gaussian fit is presented for this line.}
    \label{fig:h2o}
\end{figure*}

In addition to the above, we identified a possible detection of H$_2$O at rest frequency of 906.205 GHz or HCN(10-9) at a rest frequency of 906.24 GHz (or a combination of the two, given the coarse spectral resolution). This tentative detection is shown in Fig. \ref{fig:h2o}. 

We also carried out an independent, blind search using the Source Finding Application (SoFiA\footnote{\url{https://github.com/SoFiA-Admin/SoFiA}}; \citealt{serra15}). SoFiA provides a selection of emission line-search in data cubes, as well as routines to smooth, spatially and spectrally, the data cubes.
Prior to running SoFiA, we removed continuum emission in the cubes either by directly subtracting the continuum image or by removing the median value estimated in the cubes along the frequency axis. We adopted Gaussian kernels to smooth the cubes spatially (widths of 3, 6, and 9\,pixels, for a pixel scale of $\sim$0.14\arcsec) and in velocity (3, 6, 9, 12, 15, 18\,channel smoothing, for $\sim$28\,km/s channel width). For consistency, we ran the same kernels on the inverted cubes for reliability assessment. No additional lines were found by this procedure. 

\section{Discussion}
\label{sec:discuss}

The purpose of this study was to derive the multiplicity rates around quasars with intense star formation in their hosts, as indicated by their high FIR luminosities. The work complements previous studies that were based on smaller samples and/or lower spatial resolution, in an attempt to confirm or refute the extremely high SFRs in the hosts of these quasars and to assess the role of the environment on the concomitant accretion onto SMBHs and extreme star formation. 

Of the 152 SDSS quasars followed up by ALMA in Band 7 on the framework of this project, 142 were detected at 870 \mums at or above 3$\sigma$, with 135 detected at above 5$\sigma$. This sample of submm quasars is the largest such sample ever observed with a uniform spatial resolution (0.8\arcsec). It increases the total number of SDSS quasars that have been observed with ALMA as science targets \citep[see][]{wong23} by about 40\%, and more than doubles the SDSS quasars with ALMA Band 7 detections. 

Our analysis shows that about a third of the FIR-bright quasars have secondary submm counterparts within projected distances of a few to some tens of kpc.
At the limited SDSS resolution, none of the quasars in the sample bear signs of recent merger activity (i.e. they are point-like and have no features like tidal tails) but the spacial resolution and sensitivity of SDSS are too low for a proper assessment. 
Furthermore, lacking spectroscopic information of the secondary counterparts, it is not possible to assess whether these are physical associations at the redshift of the quasar or chance associations. 

The FIR selection of the sample inherently biases it towards systems with more gas and dust, which are also more likely to be involved in mergers. In contrast, SDSS quasars selected optically, may exhibit a lower observed merger rate. This could be because they include quasars triggered by secular processes rather than mergers, or because they involve quasars in the late stages of a merger, where multiple counterparts are more difficult to resolve.

\subsection{Multiplicities and associations}
\label{sec:multi}

\begin{figure}
    \centering
    \includegraphics[width=0.5\textwidth]{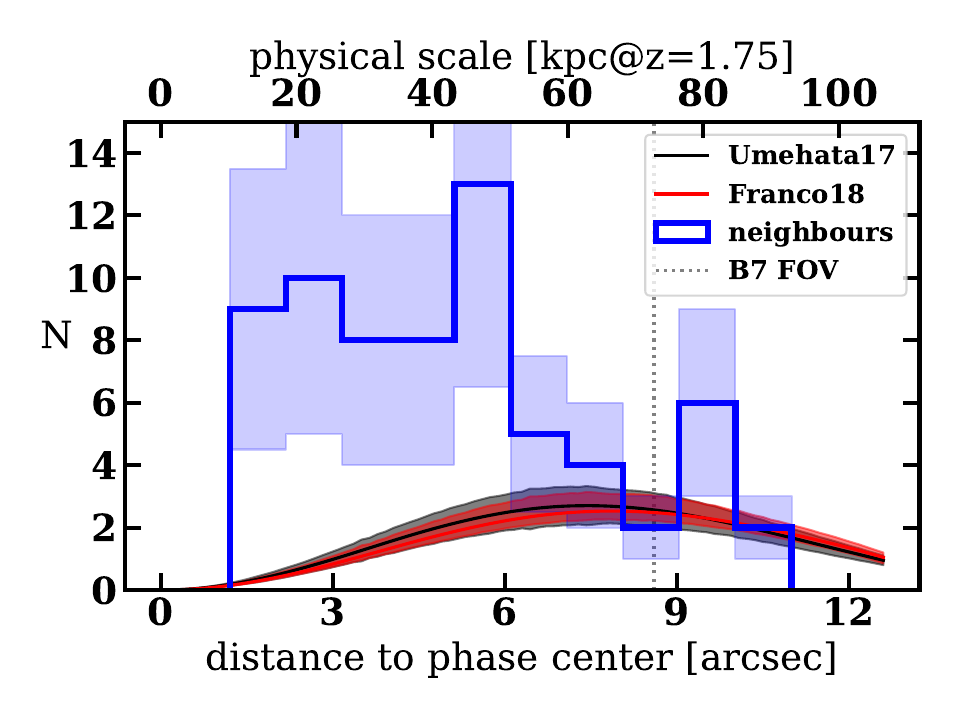}
    \caption{Comparison between the number of sources detected around the quasar sample at separations >1\arcsec\ and SNR > 5 (blue, with Poisson errors as shaded regions) and the expected field source counts from \citet[][black]{umehata17} and \citet[][red]{franco18}. Red and grey shaded areas indicate the 1$\sigma$ confidence regions from 1000 bootstrap realisations, based on the parameter uncertainties reported in those works. Counts represent the expected number of unrelated serendipitous sources over a total area corresponding to apertures of increasing radius (x-axis), accounting for primary beam attenuation—hence the drop beyond the 0.5 dilution fraction (dotted line). The top axis gives projected physical scales at the sample’s median redshift ($z = 1.75$).
    }
    \label{fig:countsvsexp}
\end{figure}

In order to assess the probability of the secondary submm counterparts found by the present analysis not being chance associations, but, instead physical pairs or triplets, we compared the number counts in the fields of the quasars with those derived from submm number counts at 1.1 mm in deep ALMA fields, namely SSA22 \citep{umehata17} and GOODS-ALMA \citep{franco18}. To directly compare  1.1 mm number counts to our 870 \mums counts, we scale the flux density of the counts as S$_{1.1 mm}$/S$_{870 \mu m}$ = 0.56, following \cite{hatsukade16} and \cite{franco18}. This value is derived assuming a modified black body model with a dust emissivity $\beta$ = 1.5 \citep[e.g][]{gordon10} and a dust temperature T = 35 K, typical for SMGs \citep{kovacs06}.
More specifically, we compared the incidence of detected companions with the expected field counts as predicted by \citet[][black line/region]{umehata17} and \citet[][red line/region]{franco18}. Overall, for the total survey area (152 fields comprising $\sim$9.8\,arcmin$^2$), \cite{umehata17} and \citet{franco18} predict 6.9$\pm$3.9 and 6.3$\pm$1.9 serendipitous (i.e. unrelated to the quasars) sources, respectively. This is a factor of ten lower than the total of 67 secondary counterparts detected at 870 \mum. The excess is particularly striking up to the distance of 6\arcsec, beyond which counts around the quasars in our sample are compatible (within the errors) with the counts from the deep fields. The comparison is shown in Fig.~\ref{fig:countsvsexp} and it is strongly indicative of physical rather than chance associations between the primary and secondary sources. 

Caution must be paid, however, in the interpretation of the multiplicities, as results based on Illustris simulations show that projected spatial separation and relative velocity of members of interactive systems are {\it both} equally important factors in determining the probability that interactive objects will merge \citep{ventou19}. Since for the multiple objects in our sample we know neither the redshift of the secondary sources nor the relative velocities, we cannot assess the probability of these systems being real mergers. Nevertheless, at least a fraction of the secondary counterparts must be SMGs, undergoing intense star formation, located at distances indicative of interacting pairs.  

Quasar-quasar or quasar-SMG pairs at high redshift can be used as suitable tracers of the large–scale distribution of galaxies and as proxies for over-dense regions (\citealt{herwig25} but see also \citealt{chen23}, as caution must be paid to the interpretation of over-densities), especially given their high luminosity and dedicated surveys. Multiple studies \citep[e.g.][]{wylezalek13, silva15, banerji21, izumi24}  have shown that $z>1.5$ quasars are often found in overdense environments rich in dusty, star-forming galaxies, suggesting that mergers are one of the mechanisms for synchronising rapid SMBH growth and star formation in quasar hosts.

Galaxy mergers, particularly major mergers, are known to be linked to quasar activity, but the precise relationship and its evolution across cosmic time are complex and still debated. While some studies show a higher incidence of mergers in quasar host galaxies compared to inactive galaxies, particularly for powerful radio-loud quasars \citep[e.g.][]{breiding24}, others find no significant correlation \citep{marian19} and results remain inconclusive especially at higher redshifts.

Given the above discussion, we can interpret the findings of our study as an upper limit of $\sim$30\% to the fraction of FIR- and optically-bright quasars in which a merger event could trigger concomitant accretion onto the central SMBH and intense star formation in their hosts, as well as star formation in the host of the merger partner. 

The increase of the multiplicity rate with redshift by a factor of $\sim$2.5 between redshifts 1 and 2.5 reported in Sect. \ref{sec:fluxzbal} is in line with observations of X-ray selected, unobscured AGN in the COSMOS field showing an increase in the number of AGN with disturbed morphologies (indicative of mergers) by a factor of $\sim$4 from redshift 0.5 to 2 \citep{hewlett17}. Nevertheless, the same work reports a total 15\% of AGN with morphologically disturbed hosts, a fraction considerably smaller than the 35\% of objects with multiple counterparts reported here that might be part of a merging system. In fact, the fraction of AGN that are part of merging systems at redshifts typically between 1 and 3 varies wildly in the literature even among the brightest systems, from a mere 10\% \citep{treister12} to above 70\% \citep{glikman15}.

The $\sim$60\% of FIR-bright SDSS quasars with a single submm counterpart located with 0.5\arcsec \, from the optical coordinates (see Fig. \ref{fig:histos}) is in line with the findings of the pilot project  \citep{hatzimi18} despite the coarser (by a factor of $\sim$5) spatial resolution of the data in the pilot study. Excluding projection effects (i.e. the companion lying along the line of sight and, hence, undetectable) and a post-coalescence phase where the merger would no longer be detectable, and given the absence of disturbed morphologies or tidal features (within the limitations of the data in hand), these numbers highlight that, while mergers enhance gas inflow efficiency, there must be other viable alternatives for driving synchronous SMBH growth and extreme star formation. Indeed, recent cosmological simulations support the idea of little to no connection between mergers and accretion \citep[e.g.][and references therein]{steinborn18, ricarte19}. In these simulations, SMBHs and their hosts assemble in parallel, regardless of their larger scale environment. 
Viable alternatives to mergers other than the commonly invoked bar-type instabilities, could be gas inflows via smooth accretion from cosmic filaments or halo reservoirs \citep{elbaz09}. Such processes can fuel both star formation and AGN activity without gravitational interactions, particularly in high-redshift environments where cold gas streams are prevalent \citep{zhu24}.

Mergers are also invoked to explain the BAL features seen in $\sim$20\% of optically-selected quasars, as the outflows presumably responsible for these features might occur at a young stage in a quasar's lifetime, when it has been recently (re)fuelled by a merger, that also triggers star formation in their hosts. LoBAL quasars, in particular, might be a short-lived transition between a merger-induced starburst and an optically luminous quasar \citep[e.g.][]{boroson92}. Alternatively, under an orientation scenario, BAL winds exist in all quasars in line with the AGN unified model \citep{antonucci93} but can only be viewed along particular lines of sight due to the covering factor of the BAL region \citep[e.g.][]{becker00}.  
The independence of the incidence of multiplicities on ``balnicity'' reported here, in line with studies of large samples that found BAL quasars to live in similar environments as non-BAL quasars \citep{shen08}, implies that the BAL phenomenon is likely not driven by, or associated with, differences in the larger-scale dusty or star-forming environments of quasars. Instead, it is likely driven by processes intrinsic to the quasars themselves or their immediate surroundings, supporting the view that BAL and non-BAL quasars are drawn from the same parent population, differing mainly in orientation or other internal properties, in support of the line-of-sight scenario. Note that, although the young-phase scenario described above is primarily relevant to the LoBAL quasars, the single object in our sample with characteristics compatible with a LoBAL nature does not have a secondary counterpart.

\subsection{CO transitions in the low resolution spectra}
\label{sec:co}

CO is the second most abundant molecule in the Universe and the best tracer of H$_2$ \citep{solomon05}. Intermediate transitions like CO(6-5) and CO(7-6) are indicative of the presence of warm and dense molecular gas in the host galaxy's star-forming regions. While intermediate CO transitions are seen both at low redshift \citep[e.g.][and references therein]{molyneux24} as well as at redshifts above $\sim$6 \citep[e.g.][]{bertoldi03, decarli22}, works reporting intermediate CO transitions seen in quasars at redshifts comparable to those in our sample are sparse. \cite{stacey22} and \cite{scholtz23} discuss the detection of CO(7-6) in the spectra of strongly reddened quasars, \cite{balashev25} report the detection of CO(7-6) in a quasar - SMG merging system at a redshift of 2.66, while, to our knowledge, there are no references to CO(6-5) observations in quasars at or near ``cosmic noon'' in the literature, with the two detections discussed in this paper being the first ones ever reported. 

In our sample, we detect two out of three expected CO(6–5) lines and three out of five expected CO(7–6) lines. Assuming a typical CO spectral line energy distribution (COSLED) for quasars, as reported by e.g. \citet{vanderwerf10} and \citet{carilli13}, we estimate that the sensitivity of our ALMA observations is sufficient to detect all CO transitions with J$_{upp}$ between 6 and 10 at a significance level above 6$\sigma$, with the exception of one source that lacks a submm continuum detection. The absence of these higher-J CO lines in the remaining sources thus suggests that the molecular gas is not sufficiently excited to populate these transitions. As such high-J lines require elevated gas temperatures and densities, their non-detection implies moderate excitation conditions without the presence of strong heating or extreme environments. Indeed, CO excitation models \citep[e.g.][]{meijerink07, kamenetzky16} associate high-J line emission primarily with AGN-driven processes. Our conclusions are also consistent with \citet{farrah22}, who report significant AGN contribution to CO transitions with J$_{upp}$=9 and above only. 

For transitions with J$_{upp} \geq 11$, our sensitivity would allow detections only at or below the 5$\sigma$ level, and we therefore refrain from drawing firm conclusions for these higher transitions.  
Our detection of intermediate-J transitions, combined with the lack of higher-J lines, align with findings by \cite{kirkpatrick19}, who also observed a lack of high-J CO lines in a similar redshift AGN sample. Both results suggests the molecular gas is predominantly in a cooler, less dense phase, typical of star-forming galaxies \citep[e.g.][]{shangguan20}. Overall, the gas is neither warm nor dense enough to reach the extreme conditions needed for significant high-J CO emission.

To further explore whether the mid-J CO transitions might originate from X-ray-dominated regions (XDRs) as suggested by e.g. \cite{esposito24}, and therefore correlate with the presence of X-ray emission, we searched for X-ray counterparts to our sample in the XMM and {\it Chandra} catalogues. We identified 27 matches in the 4XMM-DR14 catalogue \citep{webb2020} within 5\arcsec \, of the SDSS quasar coordinates. Based on their redshifts and the ALMA setup, seven were expected to show detectable CO lines: one CO(7-6), three CO(8-7), one CO(9-8), one CO(11-10), and one CO(12-11). Of these, only the CO(7-6) transition was detected.
The lack of higher-J line detections suggests that XDRs do not significantly impact the molecular gas in the quasar hosts. We also identified eight matches in the {\it Chandra} Source Catalogue Release 2.1 \citep{evans2024}, but none were expected to show CO emission.

Directly comparing intermediate-J CO detection rates across redshifts is challenging due to small sample sizes. \cite{decarli22} report ten out of ten  detections of CO(7–6) and six out of ten detections of CO(6–5) in their $z\sim 6$ quasar sample, broadly consistent with our findings. In contrast, \cite{salvestrini25} found no significant detections in five quasars at $z>7$, indicating that quasar hosts during the Epoch of Reionisation likely contained less cold molecular gas than those at lower redshifts. The small number of detections involved in our work as well as the above works makes firm conclusions difficult. The combined findings, however, suggest that gas excitation may vary with redshift or quasar luminosity.

H$_2$O transitions in the FIR are commonly present and with intensities comparable to CO transitions, both at the local Universe \citep[e.g.][]{vanderwerf10} and high-redshift quasar hosts \citep{vanderwerf11} and Ultra Luminous IR Galaxies \citep{omont13}. With data from the SPIRE Fourier Transform Spectrometer onboard {\it Herschel}, \cite{yang13} showed the luminosity of submm H$_2$O transitions to correlate with the infrared luminosity over three orders of magnitude, and suggested that these transitions trace star formation. Our own tentative detection of H$_2$O in the host of a FIR-bright quasar comes as yet another indication in favour of the above suggestion.

\subsection{On the extreme star formation rates}
\label{sec:next}

More than half the $z>1$ quasars in the parent sample of \cite{pitchford16}, and  56\% of the quasars in our sample exhibit FIR-derived SFRs exceeding 1000 \Msun yr$^{-1}$. 
The increase of L$_{\rm IR}$ and SFRs with multiplicity strongly suggests that {\it Herschel}-derived FIR properties have to be carefully assessed, as at least a fraction of the extreme FIR-based SFRs may be due of the contribution of multiple sources to the FIR fluxes.

We address this issue in a follow up paper where the SEDs of the 152 quasars are going to be reanalysed (Hatziminaoglou et al., in preparation). In this up-coming work, the 870 \mums data point is included to the optical-to-FIR SEDs of the objects. For sources with multiple submm counterparts within 10\arcsec \, of the quasar’s optical position, FIR fluxes are weighted according to the their relative 870 \mums fluxes. The SED fitting procedure described in \cite{pitchford16} is applied again to derive L$_{\rm IR}$ and SFRs, enabling a direct comparison with our previous findings. This allows for a reassessment of the extreme SFRs in quasar hosts, as well as an analysis of dust temperatures in quasars at redshifts between 1 and 4.

\begin{acknowledgements}
    EH acknowledges support from the Fundación Occident and the Instituto de Astrofísica de Canarias under the Visiting Researcher Programme 2022-2024 agreed between both institutions. RS thanks the Coordenação de Aperfeiçoamento de Pessoal de Nível Superior (CAPES—Brasil) and the CAPES-Print program for funding their stay at ESO, as well as ESO for their hospitality in the period April to December 2023 during which part of the work was carried out. A.F. acknowledges support from project "VLT- MOONS" CRAM 1.05.03.07, INAF Large Grant 2022 "The metal circle: a new sharp view of the baryon cycle up to Cosmic Dawn with the latest generation IFU facilities" and INAF Large Grant 2022 "Dual and binary SMBH in the multi-messenger era”. AB acknowledges the support of the EU-ARC.CZ Large Research Infrastructure grant project LM2023059 of the Ministry of Education, Youth and Sports of the Czech Republic. This paper makes use of the following ALMA data: ADS/JAO.ALMA\#2022.1.00029.S. ALMA is a partnership of ESO (representing its member states), NSF (USA) and NINS (Japan), together with NRC (Canada), NSTC and ASIAA (Taiwan), and KASI (Republic of Korea), in cooperation with the Republic of Chile. The Joint ALMA Observatory is operated by ESO, AUI/NRAO and NAOJ. This work makes use of TOPCAT \citep{taylor05}. 
    Funding for the Sloan Digital Sky Survey IV has been provided by the  Alfred P. Sloan Foundation, the U.S. Department of Energy Office of Science, and the Participating Institutions. SDSS-IV acknowledges support and resources from the Center for High Performance Computing  at the University of Utah. The SDSS website is www.sdss4.org. SDSS-IV is managed by the Astrophysical Research Consortium for the Participating Institutions of the SDSS Collaboration including the Brazilian Participation Group, the Carnegie Institution for Science, Carnegie Mellon University, Center for Astrophysics | Harvard \& Smithsonian, the Chilean Participation Group, the French Participation Group, Instituto de Astrof\'isica de Canarias, The Johns Hopkins University, Kavli Institute for the Physics and Mathematics of the Universe (IPMU) / University of Tokyo, the Korean Participation Group, Lawrence Berkeley National Laboratory, Leibniz Institut f\"ur Astrophysik Potsdam (AIP),  Max-Planck-Institut f\"ur Astronomie (MPIA Heidelberg), Max-Planck-Institut f\"ur Astrophysik (MPA Garching), Max-Planck-Institut f\"ur Extraterrestrische Physik (MPE), National Astronomical Observatories of China, New Mexico State University, New York University, University of Notre Dame, Observat\'ario Nacional / MCTI, The Ohio State University, Pennsylvania State University, Shanghai Astronomical Observatory, United Kingdom Participation Group, Universidad Nacional Aut\'onoma de M\'exico, University of Arizona, University of Colorado Boulder, University of Oxford, University of Portsmouth, University of Utah, University of Virginia, University of Washington, University of Wisconsin, Vanderbilt University, and Yale University.
\end{acknowledgements}

\bibliographystyle{aa}
\bibliography{ALMASDSSquasars.bib}

\begin{thebibliography}{95}
\expandafter\ifx\csname natexlab\endcsname\relax\def\natexlab#1{#1}\fi

\bibitem[{{Antonucci}(1993)}]{antonucci93}
{Antonucci}, R. 1993, \araa, 31, 473

\bibitem[{{Balashev} {et~al.}(2025){Balashev}, {Noterdaeme}, {Gupta},
  {Krogager}, {Combes}, {L{\'o}pez}, {Petitjean}, {Omont}, {Srianand}, \&
  {Cuellar}}]{balashev25}
{Balashev}, S., {Noterdaeme}, P., {Gupta}, N., {et~al.} 2025, \nat, 641, 1137

\bibitem[{{Banerji} {et~al.}(2021){Banerji}, {Jones}, {Carniani}, {DeGraf}, \&
  {Wagg}}]{banerji21}
{Banerji}, M., {Jones}, G.~C., {Carniani}, S., {DeGraf}, C., \& {Wagg}, J.
  2021, \mnras, 503, 5583

\bibitem[{{Becker} {et~al.}(2000){Becker}, {White}, {Gregg}, {Brotherton},
  {Laurent-Muehleisen}, \& {Arav}}]{becker00}
{Becker}, R.~H., {White}, R.~L., {Gregg}, M.~D., {et~al.} 2000, \apj, 538, 72

\bibitem[{{Bertoldi} {et~al.}(2003){Bertoldi}, {Cox}, {Neri}, {Carilli},
  {Walter}, {Omont}, {Beelen}, {Henkel}, {Fan}, {Strauss}, \&
  {Menten}}]{bertoldi03}
{Bertoldi}, F., {Cox}, P., {Neri}, R., {et~al.} 2003, \aap, 409, L47

\bibitem[{{Boroson} \& {Meyers}(1992)}]{boroson92}
{Boroson}, T.~A. \& {Meyers}, K.~A. 1992, \apj, 397, 442

\bibitem[{{Breiding} {et~al.}(2024){Breiding}, {Chiaberge}, {Lambrides},
  {Meyer}, {Willner}, {Hilbert}, {Haas}, {Miley}, {Perlman}, {Barthel},
  {O'Dea}, {Capetti}, {Wilkes}, {Baum}, {Macchetto}, {Sparks}, {Tremblay}, \&
  {Norman}}]{breiding24}
{Breiding}, P., {Chiaberge}, M., {Lambrides}, E., {et~al.} 2024, \apj, 963, 91

\bibitem[{{Bussmann} {et~al.}(2015){Bussmann}, {Riechers}, {Fialkov},
  {Scudder}, {Hayward}, {Cowley}, {Bock}, {Calanog}, {Chapman}, {Cooray}, {De
  Bernardis}, {Farrah}, {Fu}, {Gavazzi}, {Hopwood}, {Ivison}, {Jarvis},
  {Lacey}, {Loeb}, {Oliver}, {P{\'e}rez-Fournon}, {Rigopoulou}, {Roseboom},
  {Scott}, {Smith}, {Vieira}, {Wang}, \& {Wardlow}}]{bussmann15}
{Bussmann}, R.~S., {Riechers}, D., {Fialkov}, A., {et~al.} 2015, \apj, 812, 43

\bibitem[{{Cao Orjales} {et~al.}(2012){Cao Orjales}, {Stevens}, {Jarvis},
  {Smith}, {Hardcastle}, {Auld}, {Baes}, {Cava}, {Clements}, {Cooray},
  {Coppin}, {Dariush}, {De Zotti}, {Dunne}, {Dye}, {Eales}, {Hopwood}, {Hoyos},
  {Ibar}, {Ivison}, {Maddox}, {Page}, \& {Valiante}}]{caoorjales12}
{Cao Orjales}, J.~M., {Stevens}, J.~A., {Jarvis}, M.~J., {et~al.} 2012, \mnras,
  427, 1209

\bibitem[{{Carilli} \& {Walter}(2013)}]{carilli13}
{Carilli}, C.~L. \& {Walter}, F. 2013, \araa, 51, 105

\bibitem[{{Chen} {et~al.}(2023){Chen}, {Ivison}, {Zwaan}, {Klitsch},
  {P{\'e}roux}, {Lovell}, {Lagos}, {Biggs}, \& {Bollo}}]{chen23}
{Chen}, J., {Ivison}, R.~J., {Zwaan}, M.~A., {et~al.} 2023, \aap, 675, L10

\bibitem[{{Cowley} {et~al.}(2015){Cowley}, {Lacey}, {Baugh}, \&
  {Cole}}]{cowley15}
{Cowley}, W.~I., {Lacey}, C.~G., {Baugh}, C.~M., \& {Cole}, S. 2015, \mnras,
  446, 1784

\bibitem[{{Decarli} {et~al.}(2022){Decarli}, {Pensabene}, {Venemans}, {Walter},
  {Ba{\~n}ados}, {Bertoldi}, {Carilli}, {Cox}, {Fan}, {Farina}, {Ferkinhoff},
  {Groves}, {Li}, {Mazzucchelli}, {Neri}, {Riechers}, {Uzgil}, {Wang}, {Wang},
  {Weiss}, {Winters}, \& {Yang}}]{decarli22}
{Decarli}, R., {Pensabene}, A., {Venemans}, B., {et~al.} 2022, \aap, 662, A60

\bibitem[{{Di Matteo} {et~al.}(2005){Di Matteo}, {Springel}, \&
  {Hernquist}}]{dimatteo05}
{Di Matteo}, T., {Springel}, V., \& {Hernquist}, L. 2005, \nat, 433, 604

\bibitem[{{Draper} \& {Ballantyne}(2012)}]{draper12}
{Draper}, A.~R. \& {Ballantyne}, D.~R. 2012, \apj, 751, 72

\bibitem[{{Drouart} {et~al.}(2014){Drouart}, {De Breuck}, {Vernet}, {Seymour},
  {Lehnert}, {Barthel}, {Bauer}, {Ibar}, {Galametz}, {Haas}, {Hatch},
  {Mullaney}, {Nesvadba}, {Rocca-Volmerange}, {R{\"o}ttgering}, {Stern}, \&
  {Wylezalek}}]{drouart14}
{Drouart}, G., {De Breuck}, C., {Vernet}, J., {et~al.} 2014, \aap, 566, A53

\bibitem[{{Elbaz} {et~al.}(2009){Elbaz}, {Jahnke}, {Pantin}, {Le Borgne}, \&
  {Letawe}}]{elbaz09}
{Elbaz}, D., {Jahnke}, K., {Pantin}, E., {Le Borgne}, D., \& {Letawe}, G. 2009,
  \aap, 507, 1359

\bibitem[{{Esposito} {et~al.}(2024){Esposito}, {Vallini}, {Pozzi}, {Casasola},
  {Alonso-Herrero}, {Garc{\'\i}a-Burillo}, {Decarli}, {Calura}, {Vignali},
  {Mingozzi}, {Gruppioni}, \& {Sengupta}}]{esposito24}
{Esposito}, F., {Vallini}, L., {Pozzi}, F., {et~al.} 2024, \mnras, 527, 8727

\bibitem[{{Evans} {et~al.}(2024){Evans}, {Evans}, {Mart{\'\i}nez-Galarza},
  {Miller}, {Primini}, {Azadi}, {Burke}, {Civano}, {D'Abrusco}, {Fabbiano},
  {Graessle}, {Grier}, {Houck}, {Lauer}, {McCollough}, {Nowak}, {Plummer},
  {Rots}, {Siemiginowska}, \& {Tibbetts}}]{evans2024}
{Evans}, I.~N., {Evans}, J.~D., {Mart{\'\i}nez-Galarza}, J.~R., {et~al.} 2024,
  \apjs, 274, 22

\bibitem[{{Farrah} {et~al.}(2003){Farrah}, {Afonso}, {Efstathiou},
  {Rowan-Robinson}, {Fox}, \& {Clements}}]{farrah03}
{Farrah}, D., {Afonso}, J., {Efstathiou}, A., {et~al.} 2003, \mnras, 343, 585

\bibitem[{{Farrah} {et~al.}(2022){Farrah}, {Efstathiou}, {Afonso}, {Clements},
  {Croker}, {Hatziminaoglou}, {Joyce}, {Lebouteiller}, {Lee}, {Lonsdale},
  {Pearson}, {Petty}, {Pitchford}, {Rigopoulou}, {Verma}, \& {Wang}}]{farrah22}
{Farrah}, D., {Efstathiou}, A., {Afonso}, J., {et~al.} 2022, Universe, 9, 3

\bibitem[{{Franco} {et~al.}(2018){Franco}, {Elbaz}, {B{\'e}thermin},
  {Magnelli}, {Schreiber}, {Ciesla}, {Dickinson}, {Nagar}, {Silverman},
  {Daddi}, {Alexander}, {Wang}, {Pannella}, {Le Floc'h}, {Pope}, {Giavalisco},
  {Maury}, {Bournaud}, {Chary}, {Demarco}, {Ferguson}, {Finkelstein}, {Inami},
  {Iono}, {Juneau}, {Lagache}, {Leiton}, {Lin}, {Magdis}, {Messias},
  {Motohara}, {Mullaney}, {Okumura}, {Papovich}, {Pforr}, {Rujopakarn},
  {Sargent}, {Shu}, \& {Zhou}}]{franco18}
{Franco}, M., {Elbaz}, D., {B{\'e}thermin}, M., {et~al.} 2018, \aap, 620, A152

\bibitem[{{Glikman} {et~al.}(2015){Glikman}, {Simmons}, {Mailly}, {Schawinski},
  {Urry}, \& {Lacy}}]{glikman15}
{Glikman}, E., {Simmons}, B., {Mailly}, M., {et~al.} 2015, \apj, 806, 218

\bibitem[{{Gordon} {et~al.}(2010){Gordon}, {Galliano}, {Hony}, {Bernard},
  {Bolatto}, {Bot}, {Engelbracht}, {Hughes}, {Israel}, {Kemper}, {Kim}, {Li},
  {Madden}, {Matsuura}, {Meixner}, {Misselt}, {Okumura}, {Panuzzo}, {Rubio},
  {Reach}, {Roman-Duval}, {Sauvage}, {Skibba}, \& {Tielens}}]{gordon10}
{Gordon}, K.~D., {Galliano}, F., {Hony}, S., {et~al.} 2010, \aap, 518, L89

\bibitem[{{Griffin} {et~al.}(2010){Griffin}, {Abergel}, {Abreu}, {Ade},
  {Andr{\'e}}, {Augueres}, {Babbedge}, {Bae}, {Baillie}, {Baluteau}, {Barlow},
  {Bendo}, {Benielli}, {Bock}, {Bonhomme}, {Brisbin}, {Brockley-Blatt},
  {Caldwell}, {Cara}, {Castro-Rodriguez}, {Cerulli}, {Chanial}, {Chen},
  {Clark}, {Clements}, {Clerc}, {Coker}, {Communal}, {Conversi}, {Cox},
  {Crumb}, {Cunningham}, {Daly}, {Davis}, {de Antoni}, {Delderfield}, {Devin},
  {di Giorgio}, {Didschuns}, {Dohlen}, {Donati}, {Dowell}, {Dowell}, {Duband},
  {Dumaye}, {Emery}, {Ferlet}, {Ferrand}, {Fontignie}, {Fox}, {Franceschini},
  {Frerking}, {Fulton}, {Garcia}, {Gastaud}, {Gear}, {Glenn}, {Goizel},
  {Griffin}, {Grundy}, {Guest}, {Guillemet}, {Hargrave}, {Harwit}, {Hastings},
  {Hatziminaoglou}, {Herman}, {Hinde}, {Hristov}, {Huang}, {Imhof}, {Isaak},
  {Israelsson}, {Ivison}, {Jennings}, {Kiernan}, {King}, {Lange}, {Latter},
  {Laurent}, {Laurent}, {Leeks}, {Lellouch}, {Levenson}, {Li}, {Li},
  {Lilienthal}, {Lim}, {Liu}, {Lu}, {Madden}, {Mainetti}, {Marliani}, {McKay},
  {Mercier}, {Molinari}, {Morris}, {Moseley}, {Mulder}, {Mur}, {Naylor},
  {Nguyen}, {O'Halloran}, {Oliver}, {Olofsson}, {Olofsson}, {Orfei}, {Page},
  {Pain}, {Panuzzo}, {Papageorgiou}, {Parks}, {Parr-Burman}, {Pearce},
  {Pearson}, {P{\'e}rez-Fournon}, {Pinsard}, {Pisano}, {Podosek}, {Pohlen},
  {Polehampton}, {Pouliquen}, {Rigopoulou}, {Rizzo}, {Roseboom}, {Roussel},
  {Rowan-Robinson}, {Rownd}, {Saraceno}, {Sauvage}, {Savage}, {Savini},
  {Sawyer}, {Scharmberg}, {Schmitt}, {Schneider}, {Schulz}, {Schwartz},
  {Shafer}, {Shupe}, {Sibthorpe}, {Sidher}, {Smith}, {Smith}, {Smith},
  {Spencer}, {Stobie}, {Sudiwala}, {Sukhatme}, {Surace}, {Stevens}, {Swinyard},
  {Trichas}, {Tourette}, {Triou}, {Tseng}, {Tucker}, {Turner}, {Vaccari},
  {Valtchanov}, {Vigroux}, {Virique}, {Voellmer}, {Walker}, {Ward}, {Waskett},
  {Weilert}, {Wesson}, {White}, {Whitehouse}, {Wilson}, {Winter}, {Woodcraft},
  {Wright}, {Xu}, {Zavagno}, {Zemcov}, {Zhang}, \& {Zonca}}]{griffin10}
{Griffin}, M.~J., {Abergel}, A., {Abreu}, A., {et~al.} 2010, \aap, 518, L3

\bibitem[{{Harris} {et~al.}(2016){Harris}, {Farrah}, {Schulz},
  {Hatziminaoglou}, {Viero}, {Anderson}, {B{\'e}thermin}, {Chapman},
  {Clements}, {Cooray}, {Efstathiou}, {Feltre}, {Hurley}, {Ibar}, {Lacy},
  {Oliver}, {Page}, {P{\'e}rez-Fournon}, {Petty}, {Pitchford}, {Rigopoulou},
  {Scott}, {Symeonidis}, {Vieira}, \& {Wang}}]{harris16}
{Harris}, K., {Farrah}, D., {Schulz}, B., {et~al.} 2016, \mnras, 457, 4179

\bibitem[{{Harrison} {et~al.}(2012){Harrison}, {Alexander}, {Mullaney},
  {Altieri}, {Coia}, {Charmandaris}, {Daddi}, {Dannerbauer}, {Dasyra}, {Del
  Moro}, {Dickinson}, {Hickox}, {Ivison}, {Kartaltepe}, {Le Floc'h}, {Leiton},
  {Magnelli}, {Popesso}, {Rovilos}, {Rosario}, \& {Swinbank}}]{harrison12}
{Harrison}, C.~M., {Alexander}, D.~M., {Mullaney}, J.~R., {et~al.} 2012, \apjl,
  760, L15

\bibitem[{{Hatsukade} {et~al.}(2016){Hatsukade}, {Kohno}, {Umehata},
  {Aretxaga}, {Caputi}, {Dunlop}, {Ikarashi}, {Iono}, {Ivison}, {Lee},
  {Makiya}, {Matsuda}, {Motohara}, {Nakanishi}, {Ohta}, {Tadaki}, {Tamura},
  {Wang}, {Wilson}, {Yamaguchi}, \& {Yun}}]{hatsukade16}
{Hatsukade}, B., {Kohno}, K., {Umehata}, H., {et~al.} 2016, \pasj, 68, 36

\bibitem[{{Hatziminaoglou} {et~al.}(2018){Hatziminaoglou}, {Farrah},
  {Humphreys}, {Manrique}, {P{\'e}rez-Fournon}, {Pitchford},
  {Salvador-Sol{\'e}}, \& {Wang}}]{hatzimi18}
{Hatziminaoglou}, E., {Farrah}, D., {Humphreys}, E., {et~al.} 2018, \mnras,
  480, 4974

\bibitem[{{Hatziminaoglou} {et~al.}(2010){Hatziminaoglou}, {Omont}, {Stevens},
  {Amblard}, {Arumugam}, {Auld}, {Aussel}, {Babbedge}, {Blain}, {Bock},
  {Boselli}, {Buat}, {Burgarella}, {Castro-Rodr{\'\i}guez}, {Cava}, {Chanial},
  {Clements}, {Conley}, {Conversi}, {Cooray}, {Dowell}, {Dwek}, {Dye}, {Eales},
  {Elbaz}, {Farrah}, {Fox}, {Franceschini}, {Gear}, {Glenn}, {Gonz{\'a}lez
  Solares}, {Griffin}, {Halpern}, {Ibar}, {Isaak}, {Ivison}, {Lagache},
  {Levenson}, {Lu}, {Madden}, {Maffei}, {Mainetti}, {Marchetti}, {Mortier},
  {Nguyen}, {O'Halloran}, {Oliver}, {Page}, {Panuzzo}, {Papageorgiou},
  {Pearson}, {P{\'e}rez-Fournon}, {Pohlen}, {Rawlings}, {Rigopoulou}, {Rizzo},
  {Roseboom}, {Rowan-Robinson}, {Sanchez Portal}, {Schulz}, {Scott}, {Seymour},
  {Shupe}, {Smith}, {Symeonidis}, {Trichas}, {Tugwell}, {Vaccari},
  {Valtchanov}, {Vigroux}, {Wang}, {Ward}, {Wright}, {Xu}, \&
  {Zemcov}}]{hatzimi10}
{Hatziminaoglou}, E., {Omont}, A., {Stevens}, J.~A., {et~al.} 2010, \aap, 518,
  L33

\bibitem[{{Hayward} {et~al.}(2013){Hayward}, {Behroozi}, {Somerville},
  {Primack}, {Moreno}, \& {Wechsler}}]{hayward13}
{Hayward}, C.~C., {Behroozi}, P.~S., {Somerville}, R.~S., {et~al.} 2013,
  \mnras, 434, 2572

\bibitem[{{Hern{\'a}n-Caballero} {et~al.}(2009){Hern{\'a}n-Caballero},
  {P{\'e}rez-Fournon}, {Hatziminaoglou}, {Afonso-Luis}, {Rowan-Robinson},
  {Rigopoulou}, {Farrah}, {Lonsdale}, {Babbedge}, {Clements}, {Serjeant},
  {Pozzi}, {Vaccari}, {Montenegro-Montes}, {Valtchanov},
  {Gonz{\'a}lez-Solares}, {Oliver}, {Shupe}, {Gruppioni}, {Vila-Vilar{\'o}},
  {Lari}, \& {La Franca}}]{hernan09}
{Hern{\'a}n-Caballero}, A., {P{\'e}rez-Fournon}, I., {Hatziminaoglou}, E.,
  {et~al.} 2009, \mnras, 395, 1695

\bibitem[{{Herwig} {et~al.}(2025){Herwig}, {Arrigoni Battaia}, {Chen},
  {Obreja}, {Nowotka}, {Remus}, \& {Yajima}}]{herwig25}
{Herwig}, E., {Arrigoni Battaia}, F., {Chen}, C.-C., {et~al.} 2025, arXiv
  e-prints, arXiv:2506.11193

\bibitem[{{Hewett} {et~al.}(2006){Hewett}, {Warren}, {Leggett}, \&
  {Hodgkin}}]{hewett06}
{Hewett}, P.~C., {Warren}, S.~J., {Leggett}, S.~K., \& {Hodgkin}, S.~T. 2006,
  \mnras, 367, 454

\bibitem[{{Hewlett} {et~al.}(2017){Hewlett}, {Villforth}, {Wild},
  {Mendez-Abreu}, {Pawlik}, \& {Rowlands}}]{hewlett17}
{Hewlett}, T., {Villforth}, C., {Wild}, V., {et~al.} 2017, \mnras, 470, 755

\bibitem[{{Hodge} {et~al.}(2013){Hodge}, {Karim}, {Smail}, {Swinbank},
  {Walter}, {Biggs}, {Ivison}, {Weiss}, {Alexander}, {Bertoldi}, {Brandt},
  {Chapman}, {Coppin}, {Cox}, {Danielson}, {Dannerbauer}, {De Breuck},
  {Decarli}, {Edge}, {Greve}, {Knudsen}, {Menten}, {Rix}, {Schinnerer},
  {Simpson}, {Wardlow}, \& {van der Werf}}]{hodge13}
{Hodge}, J.~A., {Karim}, A., {Smail}, I., {et~al.} 2013, \apj, 768, 91

\bibitem[{{Hopkins} {et~al.}(2006{\natexlab{a}}){Hopkins}, {Hernquist}, {Cox},
  {Di Matteo}, {Robertson}, \& {Springel}}]{hopkins06a}
{Hopkins}, P.~F., {Hernquist}, L., {Cox}, T.~J., {et~al.} 2006{\natexlab{a}},
  \apjs, 163, 1

\bibitem[{{Hopkins} {et~al.}(2006{\natexlab{b}}){Hopkins}, {Somerville},
  {Hernquist}, {Cox}, {Robertson}, \& {Li}}]{hopkins06b}
{Hopkins}, P.~F., {Somerville}, R.~S., {Hernquist}, L., {et~al.}
  2006{\natexlab{b}}, \apj, 652, 864

\bibitem[{{Hurley} {et~al.}(2017){Hurley}, {Oliver}, {Betancourt}, {Clarke},
  {Cowley}, {Duivenvoorden}, {Farrah}, {Griffin}, {Lacey}, {Le Floc'h},
  {Papadopoulos}, {Sargent}, {Scudder}, {Vaccari}, {Valtchanov}, \&
  {Wang}}]{hurley17}
{Hurley}, P.~D., {Oliver}, S., {Betancourt}, M., {et~al.} 2017, \mnras, 464,
  885

\bibitem[{Izumi {et~al.}(2024)Izumi, Matsuoka, Onoue, Strauss, Umehata,
  Silverman, Nagao, Imanishi, Kohno, Toba, Iwasawa, Nakanishi, Sawamura,
  Fujimoto, Kikuta, Kawaguchi, Aoki, \& Goto}]{izumi24}
Izumi, T., Matsuoka, Y., Onoue, M., {et~al.} 2024, \apj, 972, 116

\bibitem[{{Kamenetzky} {et~al.}(2016){Kamenetzky}, {Rangwala}, {Glenn},
  {Maloney}, \& {Conley}}]{kamenetzky16}
{Kamenetzky}, J., {Rangwala}, N., {Glenn}, J., {Maloney}, P.~R., \& {Conley},
  A. 2016, \apj, 829, 93

\bibitem[{{Kirkpatrick} {et~al.}(2019){Kirkpatrick}, {Sharon}, {Keller}, \&
  {Pope}}]{kirkpatrick19}
{Kirkpatrick}, A., {Sharon}, C., {Keller}, E., \& {Pope}, A. 2019, \apj, 879,
  41

\bibitem[{{Kirkpatrick} {et~al.}(2020){Kirkpatrick}, {Urry}, {Brewster},
  {Cooke}, {Estrada}, {Glikman}, {Hamblin}, {Ananna}, {Carlile}, {Coleman},
  {Johnson}, {Kartaltepe}, {LaMassa}, {Marchesi}, {Powell}, {Sanders},
  {Treister}, \& {Jan Turner}}]{kirkpatrick20}
{Kirkpatrick}, A., {Urry}, C.~M., {Brewster}, J., {et~al.} 2020, \apj, 900, 5

\bibitem[{{Kov{\'a}cs} {et~al.}(2006){Kov{\'a}cs}, {Chapman}, {Dowell},
  {Blain}, {Ivison}, {Smail}, \& {Phillips}}]{kovacs06}
{Kov{\'a}cs}, A., {Chapman}, S.~C., {Dowell}, C.~D., {et~al.} 2006, \apj, 650,
  592

\bibitem[{{Lamperti} {et~al.}(2021){Lamperti}, {Harrison}, {Mainieri},
  {Kakkad}, {Perna}, {Circosta}, {Scholtz}, {Carniani}, {Cicone}, {Alexander},
  {Bischetti}, {Calistro Rivera}, {Chen}, {Cresci}, {Feruglio}, {Fiore},
  {Mannucci}, {Marconi}, {Mart{\'\i}nez-Ram{\'\i}rez}, {Netzer}, {Piconcelli},
  {Puglisi}, {Rosario}, {Schramm}, {Vietri}, {Vignali}, \&
  {Zappacosta}}]{lamperti21}
{Lamperti}, I., {Harrison}, C.~M., {Mainieri}, V., {et~al.} 2021, \aap, 654,
  A90

\bibitem[{{Lutz} {et~al.}(2008){Lutz}, {Sturm}, {Tacconi}, {Valiante},
  {Schweitzer}, {Netzer}, {Maiolino}, {Andreani}, {Shemmer}, \&
  {Veilleux}}]{lutz08}
{Lutz}, D., {Sturm}, E., {Tacconi}, L.~J., {et~al.} 2008, \apj, 684, 853

\bibitem[{{Lyke} {et~al.}(2020){Lyke}, {Higley}, {McLane}, {Schurhammer},
  {Myers}, {Ross}, {Dawson}, {Chabanier}, {Martini}, {Busca}, {Mas des
  Bourboux}, {Salvato}, {Streblyanska}, {Zarrouk}, {Burtin}, {Anderson},
  {Bautista}, {Bizyaev}, {Brandt}, {Brinkmann}, {Brownstein}, {Comparat},
  {Green}, {de la Macorra}, {Mu{\~n}oz Guti{\'e}rrez}, {Hou}, {Newman},
  {Palanque-Delabrouille}, {P{\^a}ris}, {Percival}, {Petitjean}, {Rich},
  {Rossi}, {Schneider}, {Smith}, {Vivek}, \& {Weaver}}]{lyke20}
{Lyke}, B.~W., {Higley}, A.~N., {McLane}, J.~N., {et~al.} 2020, \apjs, 250, 8

\bibitem[{{Maddox} {et~al.}(2017){Maddox}, {Jarvis}, {Banerji}, {Hewett},
  {Bourne}, {Dunne}, {Dye}, {Eales}, {Furlanetto}, {Maddox}, {Smith}, \&
  {Valiante}}]{maddox17}
{Maddox}, N., {Jarvis}, M.~J., {Banerji}, M., {et~al.} 2017, \mnras, 470, 2314

\bibitem[{{Magliocchetti}(2022)}]{magliocchetti22}
{Magliocchetti}, M. 2022, \aapr, 30, 6

\bibitem[{{Marian} {et~al.}(2019){Marian}, {Jahnke}, {Mechtley}, {Cohen},
  {Husemann}, {Jones}, {Koekemoer}, {Schulze}, {van der Wel}, {Villforth}, \&
  {Windhorst}}]{marian19}
{Marian}, V., {Jahnke}, K., {Mechtley}, M., {et~al.} 2019, \apj, 882, 141

\bibitem[{{McMahon} {et~al.}(2021){McMahon}, {Banerji}, {Gonzalez}, {Koposov},
  {Bejar}, {Lodieu}, {Rebolo}, \& {VHS Collaboration}}]{mcmahon21}
{McMahon}, R.~G., {Banerji}, M., {Gonzalez}, E., {et~al.} 2021, VizieR Online
  Data Catalog, II/367

\bibitem[{{Meijerink} {et~al.}(2007){Meijerink}, {Spaans}, \&
  {Israel}}]{meijerink07}
{Meijerink}, R., {Spaans}, M., \& {Israel}, F.~P. 2007, \aap, 461, 793

\bibitem[{{Molyneux} {et~al.}(2024){Molyneux}, {Calistro Rivera}, {De Breuck},
  {Harrison}, {Mainieri}, {Lundgren}, {Kakkad}, {Circosta}, {Girdhar}, {Costa},
  {Mullaney}, {Kharb}, {Arrigoni Battaia}, {Farina}, {Alexander}, {Ward},
  {Silpa}, \& {Smit}}]{molyneux24}
{Molyneux}, S.~J., {Calistro Rivera}, G., {De Breuck}, C., {et~al.} 2024,
  \mnras, 527, 4420

\bibitem[{{Morrissey} {et~al.}(2007){Morrissey}, {Conrow}, {Barlow}, {Small},
  {Seibert}, {Wyder}, {Budav{\'a}ri}, {Arnouts}, {Friedman}, {Forster},
  {Martin}, {Neff}, {Schiminovich}, {Bianchi}, {Donas}, {Heckman}, {Lee},
  {Madore}, {Milliard}, {Rich}, {Szalay}, {Welsh}, \& {Yi}}]{morrisey07}
{Morrissey}, P., {Conrow}, T., {Barlow}, T.~A., {et~al.} 2007, \apjs, 173, 682

\bibitem[{{Mullaney} {et~al.}(2015){Mullaney}, {Alexander}, {Aird}, {Bernhard},
  {Daddi}, {Del Moro}, {Dickinson}, {Elbaz}, {Harrison}, {Juneau}, {Liu},
  {Pannella}, {Rosario}, {Santini}, {Sargent}, {Schreiber}, {Simpson}, \&
  {Stanley}}]{mullaney15}
{Mullaney}, J.~R., {Alexander}, D.~M., {Aird}, J., {et~al.} 2015, \mnras, 453,
  L83

\bibitem[{{Nguyen} {et~al.}(2020){Nguyen}, {Lira}, {Trakhtenbrot}, {Netzer},
  {Cicone}, {Maiolino}, \& {Shemmer}}]{nguyen20}
{Nguyen}, N.~H., {Lira}, P., {Trakhtenbrot}, B., {et~al.} 2020, \apj, 895, 74

\bibitem[{{Oliver} {et~al.}(2012){Oliver}, {Bock}, {Altieri}, {Amblard},
  {Arumugam}, {Aussel}, {Babbedge}, {Beelen}, {B{\'e}thermin}, {Blain},
  {Boselli}, {Bridge}, {Brisbin}, {Buat}, {Burgarella},
  {Castro-Rodr{\'\i}guez}, {Cava}, {Chanial}, {Cirasuolo}, {Clements},
  {Conley}, {Conversi}, {Cooray}, {Dowell}, {Dubois}, {Dwek}, {Dye}, {Eales},
  {Elbaz}, {Farrah}, {Feltre}, {Ferrero}, {Fiolet}, {Fox}, {Franceschini},
  {Gear}, {Giovannoli}, {Glenn}, {Gong}, {Gonz{\'a}lez Solares}, {Griffin},
  {Halpern}, {Harwit}, {Hatziminaoglou}, {Heinis}, {Hurley}, {Hwang}, {Hyde},
  {Ibar}, {Ilbert}, {Isaak}, {Ivison}, {Lagache}, {Le Floc'h}, {Levenson},
  {Faro}, {Lu}, {Madden}, {Maffei}, {Magdis}, {Mainetti}, {Marchetti},
  {Marsden}, {Marshall}, {Mortier}, {Nguyen}, {O'Halloran}, {Omont}, {Page},
  {Panuzzo}, {Papageorgiou}, {Patel}, {Pearson}, {P{\'e}rez-Fournon}, {Pohlen},
  {Rawlings}, {Raymond}, {Rigopoulou}, {Riguccini}, {Rizzo}, {Rodighiero},
  {Roseboom}, {Rowan-Robinson}, {S{\'a}nchez Portal}, {Schulz}, {Scott},
  {Seymour}, {Shupe}, {Smith}, {Stevens}, {Symeonidis}, {Trichas}, {Tugwell},
  {Vaccari}, {Valtchanov}, {Vieira}, {Viero}, {Vigroux}, {Wang}, {Ward},
  {Wardlow}, {Wright}, {Xu}, \& {Zemcov}}]{oliver12}
{Oliver}, S.~J., {Bock}, J., {Altieri}, B., {et~al.} 2012, \mnras, 424, 1614

\bibitem[{{Omont} {et~al.}(2013){Omont}, {Yang}, {Cox}, {Neri}, {Beelen},
  {Bussmann}, {Gavazzi}, {van der Werf}, {Riechers}, {Downes}, {Krips}, {Dye},
  {Ivison}, {Vieira}, {Wei{\ss}}, {Aguirre}, {Baes}, {Baker}, {Bertoldi},
  {Cooray}, {Dannerbauer}, {De Zotti}, {Eales}, {Fu}, {Gao}, {Gu{\'e}lin},
  {Harris}, {Jarvis}, {Lehnert}, {Leeuw}, {Lupu}, {Menten}, {Micha{\l}owski},
  {Negrello}, {Serjeant}, {Temi}, {Auld}, {Dariush}, {Dunne}, {Fritz},
  {Hopwood}, {Hoyos}, {Ibar}, {Maddox}, {Smith}, {Valiante}, {Bock},
  {Bradford}, {Glenn}, \& {Scott}}]{omont13}
{Omont}, A., {Yang}, C., {Cox}, P., {et~al.} 2013, \aap, 551, A115

\bibitem[{{Page} {et~al.}(2012){Page}, {Symeonidis}, {Vieira}, {Altieri},
  {Amblard}, {Arumugam}, {Aussel}, {Babbedge}, {Blain}, {Bock}, {Boselli},
  {Buat}, {Castro-Rodr{\'\i}guez}, {Cava}, {Chanial}, {Clements}, {Conley},
  {Conversi}, {Cooray}, {Dowell}, {Dubois}, {Dunlop}, {Dwek}, {Dye}, {Eales},
  {Elbaz}, {Farrah}, {Fox}, {Franceschini}, {Gear}, {Glenn}, {Griffin},
  {Halpern}, {Hatziminaoglou}, {Ibar}, {Isaak}, {Ivison}, {Lagache},
  {Levenson}, {Lu}, {Madden}, {Maffei}, {Mainetti}, {Marchetti}, {Nguyen},
  {O'Halloran}, {Oliver}, {Omont}, {Panuzzo}, {Papageorgiou}, {Pearson},
  {P{\'e}rez-Fournon}, {Pohlen}, {Rawlings}, {Rigopoulou}, {Riguccini},
  {Rizzo}, {Rodighiero}, {Roseboom}, {Rowan-Robinson}, {Portal}, {Schulz},
  {Scott}, {Seymour}, {Shupe}, {Smith}, {Stevens}, {Trichas}, {Tugwell},
  {Vaccari}, {Valtchanov}, {Viero}, {Vigroux}, {Wang}, {Ward}, {Wright}, {Xu},
  \& {Zemcov}}]{page12}
{Page}, M.~J., {Symeonidis}, M., {Vieira}, J.~D., {et~al.} 2012, \nat, 485, 213

\bibitem[{{P{\^a}ris} {et~al.}(2014){P{\^a}ris}, {Petitjean}, {Aubourg},
  {Ross}, {Myers}, {Streblyanska}, {Bailey}, {Hall}, {Strauss}, {Anderson},
  {Bizyaev}, {Borde}, {Brinkmann}, {Bovy}, {Brandt}, {Brewington},
  {Brownstein}, {Cook}, {Ebelke}, {Fan}, {Filiz Ak}, {Finley}, {Font-Ribera},
  {Ge}, {Hamann}, {Ho}, {Jiang}, {Kinemuchi}, {Malanushenko}, {Malanushenko},
  {Marchante}, {McGreer}, {McMahon}, {Miralda-Escud{\'e}}, {Muna},
  {Noterdaeme}, {Oravetz}, {Palanque-Delabrouille}, {Pan}, {Perez-Fournon},
  {Pieri}, {Riffel}, {Schlegel}, {Schneider}, {Simmons}, {Viel}, {Weaver},
  {Wood-Vasey}, {Y{\`e}che}, \& {York}}]{paris14}
{P{\^a}ris}, I., {Petitjean}, P., {Aubourg}, {\'E}., {et~al.} 2014, \aap, 563,
  A54

\bibitem[{{Pitchford} {et~al.}(2016){Pitchford}, {Hatziminaoglou}, {Feltre},
  {Farrah}, {Clarke}, {Harris}, {Hurley}, {Oliver}, {Page}, \&
  {Wang}}]{pitchford16}
{Pitchford}, L.~K., {Hatziminaoglou}, E., {Feltre}, A., {et~al.} 2016, \mnras,
  462, 4067

\bibitem[{{Ricarte} {et~al.}(2019){Ricarte}, {Tremmel}, {Natarajan}, \&
  {Quinn}}]{ricarte19}
{Ricarte}, A., {Tremmel}, M., {Natarajan}, P., \& {Quinn}, T. 2019, \mnras,
  489, 802

\bibitem[{{Rigopoulou} {et~al.}(2013){Rigopoulou}, {Hurley}, {Swinyard},
  {Virdee}, {Croxall}, {Hopwood}, {Lim}, {Magdis}, {Pearson}, {Pellegrini},
  {Polehampton}, \& {Smith}}]{rigopoulou13}
{Rigopoulou}, D., {Hurley}, P.~D., {Swinyard}, B.~M., {et~al.} 2013, \mnras,
  434, 2051

\bibitem[{{Salvestrini} {et~al.}(2025){Salvestrini}, {Feruglio}, {Tripodi},
  {Fontanot}, {Bischetti}, {De Lucia}, {Fiore}, {Hirschmann}, {Maio},
  {Piconcelli}, {Saccheo}, {Tortosa}, {Valiante}, {Xie}, \&
  {Zappacosta}}]{salvestrini25}
{Salvestrini}, F., {Feruglio}, C., {Tripodi}, R., {et~al.} 2025, \aap, 695, A23

\bibitem[{{Santini} {et~al.}(2012){Santini}, {Rosario}, {Shao}, {Lutz},
  {Maiolino}, {Alexander}, {Altieri}, {Andreani}, {Aussel}, {Bauer}, {Berta},
  {Bongiovanni}, {Brandt}, {Brusa}, {Cepa}, {Cimatti}, {Daddi}, {Elbaz},
  {Fontana}, {F{\"o}rster Schreiber}, {Genzel}, {Grazian}, {Le Floc'h},
  {Magnelli}, {Mainieri}, {Nordon}, {P{\'e}rez Garcia}, {Poglitsch}, {Popesso},
  {Pozzi}, {Riguccini}, {Rodighiero}, {Salvato}, {Sanchez-Portal}, {Sturm},
  {Tacconi}, {Valtchanov}, \& {Wuyts}}]{santini12}
{Santini}, P., {Rosario}, D.~J., {Shao}, L., {et~al.} 2012, \aap, 540, A109

\bibitem[{{Schneider} {et~al.}(2007){Schneider}, {Hall}, {Richards}, {Strauss},
  {Vanden Berk}, {Anderson}, {Brandt}, {Fan}, {Jester}, {Gray}, {Gunn},
  {SubbaRao}, {Thakar}, {Stoughton}, {Szalay}, {Yanny}, {York}, {Bahcall},
  {Barentine}, {Blanton}, {Brewington}, {Brinkmann}, {Brunner}, {Castander},
  {Csabai}, {Frieman}, {Fukugita}, {Harvanek}, {Hogg}, {Ivezi{\'c}}, {Kent},
  {Kleinman}, {Knapp}, {Kron}, {Krzesi{\'n}ski}, {Long}, {Lupton}, {Nitta},
  {Pier}, {Saxe}, {Shen}, {Snedden}, {Weinberg}, \& {Wu}}]{schneider07}
{Schneider}, D.~P., {Hall}, P.~B., {Richards}, G.~T., {et~al.} 2007, \aj, 134,
  102

\bibitem[{{Scholtz} {et~al.}(2023){Scholtz}, {Maiolino}, {Jones}, \&
  {Carniani}}]{scholtz23}
{Scholtz}, J., {Maiolino}, R., {Jones}, G.~C., \& {Carniani}, S. 2023, \mnras,
  519, 5246

\bibitem[{{Scudder} {et~al.}(2016){Scudder}, {Oliver}, {Hurley}, {Griffin},
  {Sargent}, {Scott}, {Wang}, \& {Wardlow}}]{scudder16}
{Scudder}, J.~M., {Oliver}, S., {Hurley}, P.~D., {et~al.} 2016, \mnras, 460,
  1119

\bibitem[{{Serra} {et~al.}(2015){Serra}, {Westmeier}, {Giese}, {Jurek},
  {Fl{\"o}er}, {Popping}, {Winkel}, {van der Hulst}, {Meyer}, {Koribalski},
  {Staveley-Smith}, \& {Courtois}}]{serra15}
{Serra}, P., {Westmeier}, T., {Giese}, N., {et~al.} 2015, \mnras, 448, 1922

\bibitem[{{Shangguan} {et~al.}(2020){Shangguan}, {Ho}, {Bauer}, {Wang}, \&
  {Treister}}]{shangguan20}
{Shangguan}, J., {Ho}, L.~C., {Bauer}, F.~E., {Wang}, R., \& {Treister}, E.
  2020, \apj, 899, 112

\bibitem[{{Shen} {et~al.}(2008){Shen}, {Strauss}, {Hall}, {Schneider}, {York},
  \& {Bahcall}}]{shen08}
{Shen}, Y., {Strauss}, M.~A., {Hall}, P.~B., {et~al.} 2008, \apj, 677, 858

\bibitem[{{Silva} {et~al.}(2015){Silva}, {Sajina}, {Lonsdale}, \&
  {Lacy}}]{silva15}
{Silva}, A., {Sajina}, A., {Lonsdale}, C., \& {Lacy}, M. 2015, \apjl, 806, L25

\bibitem[{{Solomon} \& {Vanden Bout}(2005)}]{solomon05}
{Solomon}, P.~M. \& {Vanden Bout}, P.~A. 2005, \araa, 43, 677

\bibitem[{{Stacey} {et~al.}(2022){Stacey}, {Costa}, {McKean}, {Sharon},
  {Calistro Rivera}, {Glikman}, \& {van der Werf}}]{stacey22}
{Stacey}, H.~R., {Costa}, T., {McKean}, J.~P., {et~al.} 2022, \mnras, 517, 3377

\bibitem[{{Stach} {et~al.}(2018){Stach}, {Smail}, {Swinbank}, {Simpson},
  {Geach}, {An}, {Almaini}, {Arumugam}, {Blain}, {Chapman}, {Chen},
  {Conselice}, {Cooke}, {Coppin}, {Dunlop}, {Farrah}, {Gullberg}, {Hartley},
  {Ivison}, {Maltby}, {Micha{\l}owski}, {Scott}, {Simpson}, {Thomson},
  {Wardlow}, \& {van der Werf}}]{stach18}
{Stach}, S.~M., {Smail}, I., {Swinbank}, A.~M., {et~al.} 2018, \apj, 860, 161

\bibitem[{{Steinborn} {et~al.}(2018){Steinborn}, {Hirschmann}, {Dolag},
  {Shankar}, {Juneau}, {Krumpe}, {Remus}, \& {Teklu}}]{steinborn18}
{Steinborn}, L.~K., {Hirschmann}, M., {Dolag}, K., {et~al.} 2018, \mnras, 481,
  341

\bibitem[{{Symeonidis}(2017)}]{symeonidis17}
{Symeonidis}, M. 2017, \mnras, 465, 1401

\bibitem[{{Taylor}(2005)}]{taylor05}
{Taylor}, M.~B. 2005, in Astronomical Society of the Pacific Conference Series,
  Vol. 347, Astronomical Data Analysis Software and Systems XIV, ed.
  P.~{Shopbell}, M.~{Britton}, \& R.~{Ebert}, 29

\bibitem[{{Trakhtenbrot} {et~al.}(2017){Trakhtenbrot}, {Lira}, {Netzer},
  {Cicone}, {Maiolino}, \& {Shemmer}}]{trakhtenbrot17}
{Trakhtenbrot}, B., {Lira}, P., {Netzer}, H., {et~al.} 2017, \apj, 836, 8

\bibitem[{{Treister} {et~al.}(2012){Treister}, {Schawinski}, {Urry}, \&
  {Simmons}}]{treister12}
{Treister}, E., {Schawinski}, K., {Urry}, C.~M., \& {Simmons}, B.~D. 2012,
  \apjl, 758, L39

\bibitem[{{Umehata} {et~al.}(2017){Umehata}, {Tamura}, {Kohno}, {Ivison},
  {Smail}, {Hatsukade}, {Nakanishi}, {Kato}, {Ikarashi}, {Matsuda}, {Fujimoto},
  {Iono}, {Lee}, {Steidel}, {Saito}, {Alexander}, {Yun}, \& {Kubo}}]{umehata17}
{Umehata}, H., {Tamura}, Y., {Kohno}, K., {et~al.} 2017, \apj, 835, 98

\bibitem[{{Valentino} {et~al.}(2021){Valentino}, {Daddi}, {Puglisi}, {Magdis},
  {Kokorev}, {Liu}, {Madden}, {G{\'o}mez-Guijarro}, {Lee}, {Cortzen},
  {Circosta}, {Delvecchio}, {Mullaney}, {Gao}, {Gobat}, {Aravena}, {Jin},
  {Fujimoto}, {Silverman}, \& {Dannerbauer}}]{valentino21}
{Valentino}, F., {Daddi}, E., {Puglisi}, A., {et~al.} 2021, \aap, 654, A165

\bibitem[{{van der Werf} {et~al.}(2011){van der Werf}, {Berciano Alba},
  {Spaans}, {Loenen}, {Meijerink}, {Riechers}, {Cox}, {Wei{\ss}}, \&
  {Walter}}]{vanderwerf11}
{van der Werf}, P.~P., {Berciano Alba}, A., {Spaans}, M., {et~al.} 2011, \apjl,
  741, L38

\bibitem[{{van der Werf} {et~al.}(2010){van der Werf}, {Isaak}, {Meijerink},
  {Spaans}, {Rykala}, {Fulton}, {Loenen}, {Walter}, {Wei{\ss}}, {Armus},
  {Fischer}, {Israel}, {Harris}, {Veilleux}, {Henkel}, {Savini}, {Lord},
  {Smith}, {Gonz{\'a}lez-Alfonso}, {Naylor}, {Aalto}, {Charmandaris}, {Dasyra},
  {Evans}, {Gao}, {Greve}, {G{\"u}sten}, {Kramer}, {Mart{\'\i}n-Pintado},
  {Mazzarella}, {Papadopoulos}, {Sanders}, {Spinoglio}, {Stacey}, {Vlahakis},
  {Wiedner}, \& {Xilouris}}]{vanderwerf10}
{van der Werf}, P.~P., {Isaak}, K.~G., {Meijerink}, R., {et~al.} 2010, \aap,
  518, L42

\bibitem[{{Ventou} {et~al.}(2019){Ventou}, {Contini}, {Bouch{\'e}}, {Epinat},
  {Brinchmann}, {Inami}, {Richard}, {Schroetter}, {Soucail}, {Steinmetz}, \&
  {Weilbacher}}]{ventou19}
{Ventou}, E., {Contini}, T., {Bouch{\'e}}, N., {et~al.} 2019, \aap, 631, A87

\bibitem[{{Viero} {et~al.}(2014){Viero}, {Asboth}, {Roseboom}, {Moncelsi},
  {Marsden}, {Mentuch Cooper}, {Zemcov}, {Addison}, {Baker}, {Beelen}, {Bock},
  {Bridge}, {Conley}, {Devlin}, {Dor{\'e}}, {Farrah}, {Finkelstein},
  {Font-Ribera}, {Geach}, {Gebhardt}, {Gill}, {Glenn}, {Hajian}, {Halpern},
  {Jogee}, {Kurczynski}, {Lapi}, {Negrello}, {Oliver}, {Papovich}, {Quadri},
  {Ross}, {Scott}, {Schulz}, {Somerville}, {Spergel}, {Vieira}, {Wang}, \&
  {Wechsler}}]{viero14}
{Viero}, M.~P., {Asboth}, V., {Roseboom}, I.~G., {et~al.} 2014, \apjs, 210, 22

\bibitem[{{Wang} {et~al.}(2014){Wang}, {Viero}, {Clarke}, {Bock}, {Buat},
  {Conley}, {Farrah}, {Guo}, {Heinis}, {Magdis}, {Marchetti}, {Marsden},
  {Norberg}, {Oliver}, {Page}, {Roehlly}, {Roseboom}, {Schulz}, {Smith},
  {Vaccari}, \& {Zemcov}}]{wang14}
{Wang}, L., {Viero}, M., {Clarke}, C., {et~al.} 2014, \mnras, 444, 2870

\bibitem[{{Webb} {et~al.}(2020){Webb}, {Coriat}, {Traulsen}, {Ballet}, {Motch},
  {Carrera}, {Koliopanos}, {Authier}, {de la Calle}, {Ceballos}, {Colomo},
  {Chuard}, {Freyberg}, {Garcia}, {Kolehmainen}, {Lamer}, {Lin}, {Maggi},
  {Michel}, {Page}, {Page}, {Perea-Calderon}, {Pineau}, {Rodriguez}, {Rosen},
  {Santos Lleo}, {Saxton}, {Schwope}, {Tom{\'a}s}, {Watson}, \&
  {Zakardjian}}]{webb2020}
{Webb}, N.~A., {Coriat}, M., {Traulsen}, I., {et~al.} 2020, \aap, 641, A136

\bibitem[{{White} {et~al.}(1997){White}, {Becker}, {Helfand}, \&
  {Gregg}}]{white97}
{White}, R.~L., {Becker}, R.~H., {Helfand}, D.~J., \& {Gregg}, M.~D. 1997,
  \apj, 475, 479

\bibitem[{{Wolfire} {et~al.}(2022){Wolfire}, {Vallini}, \&
  {Chevance}}]{wolfire22}
{Wolfire}, M.~G., {Vallini}, L., \& {Chevance}, M. 2022, \araa, 60, 247

\bibitem[{{Wong} {et~al.}(2023){Wong}, {Hatziminaoglou}, {Borkar}, {Popping},
  {P{\'e}rez-Fournon}, {Poidevin}, {Stoehr}, \& {Messias}}]{wong23}
{Wong}, A., {Hatziminaoglou}, E., {Borkar}, A., {et~al.} 2023, \mnras, 523, 23

\bibitem[{{Wright} {et~al.}(2010){Wright}, {Eisenhardt}, {Mainzer}, {Ressler},
  {Cutri}, {Jarrett}, {Kirkpatrick}, {Padgett}, {McMillan}, {Skrutskie},
  {Stanford}, {Cohen}, {Walker}, {Mather}, {Leisawitz}, {Gautier}, {McLean},
  {Benford}, {Lonsdale}, {Blain}, {Mendez}, {Irace}, {Duval}, {Liu}, {Royer},
  {Heinrichsen}, {Howard}, {Shannon}, {Kendall}, {Walsh}, {Larsen}, {Cardon},
  {Schick}, {Schwalm}, {Abid}, {Fabinsky}, {Naes}, \& {Tsai}}]{wright10}
{Wright}, E.~L., {Eisenhardt}, P. R.~M., {Mainzer}, A.~K., {et~al.} 2010, \aj,
  140, 1868

\bibitem[{{Wylezalek} {et~al.}(2013){Wylezalek}, {Galametz}, {Stern}, {Vernet},
  {De Breuck}, {Seymour}, {Brodwin}, {Eisenhardt}, {Gonzalez}, {Hatch},
  {Jarvis}, {Rettura}, {Stanford}, \& {Stevens}}]{wylezalek13}
{Wylezalek}, D., {Galametz}, A., {Stern}, D., {et~al.} 2013, \apj, 769, 79

\bibitem[{{Yang} {et~al.}(2013){Yang}, {Gao}, {Omont}, {Liu}, {Isaak},
  {Downes}, {van der Werf}, \& {Lu}}]{yang13}
{Yang}, C., {Gao}, Y., {Omont}, A., {et~al.} 2013, \apjl, 771, L24

\bibitem[{Zhu {et~al.}(2024)Zhu, Bakx, Ikeda, Umehata, Becker, Cain, Champagne,
  Fan, Fudamoto, Jin, Ma, Sun, Takeuchi, \& Tee}]{zhu24}
Zhu, Y., Bakx, T. J. L.~C., Ikeda, R., {et~al.} 2024, Research Notes of the
  AAS, 8, 284

\end{thebibliography}

\end{document}